\documentclass[11pt]{article}
\usepackage{enumitem}
\usepackage{pdfpages}
\usepackage{multirow}
\usepackage{array}
\usepackage{graphicx}
\usepackage{subcaption}
\usepackage{latexsym}
\usepackage{pstricks}
\usepackage{pst-node}
\usepackage{pst-tree}
\usepackage{url}
\usepackage{times}
\usepackage{helvet}
\usepackage{courier}
\usepackage{algorithmic}
\usepackage[ruled,noline,linesnumbered]{algorithm2e}
\usepackage{amsthm}
\usepackage{amssymb,amsmath}
\usepackage{mathrsfs}
\usepackage{mathtools}
\usepackage{comment}
\usepackage{framed}
\usepackage{xspace}
\usepackage{xcolor}
\usepackage[T1]{fontenc}
\theoremstyle{plain} % Standard style for theorems
\newtheorem{theorem}{Theorem}
\newtheorem{lemma}[theorem]{Lemma}
\newtheorem{corollary}[theorem]{Corollary}

\theoremstyle{definition}
\newtheorem{definition}{Definition}

\DeclareMathOperator*{\argmax}{arg\,max}

\DeclareMathOperator{\poly}{poly}

\DeclareMathOperator{\depth}{depth}
\DeclareMathOperator{\LCA}{LCA}
\DeclareMathOperator{\owner}{owner}
\usepackage[hypertexnames=false,colorlinks=true,citecolor=blue,
  linkcolor=blue,urlcolor=black]{hyperref}
\usepackage{geometry}
\makeatletter
\renewcommand*{\verbatim@font}{\sffamily}
\makeatother

\newcommand{\nil}{\mathsf{NIL}}
\newcommand{\Par}{\mathsf{par}}

\SetKw{Parallel}{in parallel}
\SetKw{Return}{return}
\SetKwInput{KwRequire}{Requires}
\SetAlgoInsideSkip{smallskip}
\SetAlFnt{\small}
\SetAlCapFnt{\small\bfseries}
\SetAlCapNameFnt{\small\bfseries}
\SetAlgoNlRelativeSize{-1}
\SetAlgoNoLine
\newcommand{\algstep}[2]{Algorithm~\ref{#1}, Step~(\ref{#2})}
\newcommand{\Cover}{\mathsf{cover}}

\newcommand{\J}{\mathsf{J}}
\newcommand{\Dis}{\mathsf{Dis}}
\newcounter{algorithm}
\renewcommand{\thealgorithm}{\arabic{algorithm}}

\newcommand{\algcaption}[1]{%
  \refstepcounter{algorithm}%
  \textbf{Algorithm \thealgorithm: #1}\par
  \smallskip
}

\makeatletter
\renewcommand*{\verbatim@font}{\sffamily}
\title{Parallel Spanning Forest and Biconnectivity for Well-Connected Graph in Sub-Logarithmic Time}
\author{
Bojun Liu\thanks{Shanghai Jiao Tong University, liunian23@sjtu.edu.cn.}
\and
S. Cliff Liu\thanks{Shanghai Jiao Tong University, cliffliu@sjtu.edu.cn.}
\and
Shuaibo Yang\thanks{Shanghai Jiao Tong University, ysb0315@sjtu.edu.cn.}
}
\date{}

\begin{document}

\clearpage\maketitle

%第一段只说CC和SF的区别，额外信息量，不要提到除了SF之外的东西，主要阐述SF的难点
%第二段最后一句改成可以输出rooted spanning forest
%diameter-sensitive改成以直径为参数

\begin{abstract}
%Computing a spanning forest is a fundamental problem in parallel graph algorithms. 
% Connectivity algorithms parameterized by the component
% diameter need only output component labels and may discard edge information. In contrast, a spanning forest must consist of original edges,
% although a virtual edge created during execution may represent an
% entire path in the input graph. Preserving and recovering this
% original-edge information within the same time and processor bounds is
% the main challenge.
We give a randomized ARBITRARY CRCW PRAM algorithm for computing a spanning forest of an \(n\)-vertex,
\(m\)-edge graph. Let \(d\) be the maximum component diameter.
With good probability, the algorithm runs in
\(O(\log d+\log\log n)\) time using \(O(m+n)\) processors and outputs
a rooted spanning forest of original edges with maximum depth
\(2^{O(\log d+\log\log n)}\).

For finding bridges, our algorithm identifies every bridge edge in
\(O(\log d+\log\log n)\) time. For biconnectivity, it outputs all
blocks in \(O(\log d+\log d'+\log\log n)\) time, where \(d'\) is the
bi-diameter. 

All these problems are fundamental in graph algorithms with numerous applications. Our results improve the best previous time bounds parameterized by the number of vertices and diameter. % while retaining \(O(m+n)\) processors and the same good-probability guarantee. 

Finally, we prove deterministic \(\Omega(\log n/\log\log n)\) lower bounds for
outputting a DFS Sequence, a weighted spanning forest, and an ear
decomposition.
\end{abstract}

\newpage

\tableofcontents

\newpage

\newpage

\section{Introduction}
\label{sec:intro}

Computing a spanning forest is a fundamental problem in parallel graph
algorithms. 
Connectivity algorithms need only output component labels and may discard edge information. In contrast, a spanning forest must consist of original edges,
although a virtual edge created during execution may represent an
entire path in the input graph. Preserving and recovering this
original-edge information within the same time and processor bounds is
the main challenge.
Sequentially, depth-first search gives linear-time algorithms for
connectivity and biconnectivity~\cite{Tarjan72,HT73}. As modern graph
datasets grow to billions of vertices and edges, even linear
sequential time can be prohibitive. This scale motivates parallel
algorithms that reduce the parallel time to sublinear bounds.

Let \(G\) be an undirected graph with \(n\) vertices, \(m\) edges, and
maximum component diameter \(d\). We work on an ARBITRARY CRCW PRAM
with \(O(m+n)\) processors. Classical PRAM algorithms compute
connected components in \(O(\log n)\) time~\cite{SV82,AS87}.
Randomized algorithms achieve this bound with linear
work~\cite{Gazit91,HZ96}. Halperin and Zwick~\cite{HZ01} also gave
\(O(\log n)\)-time, linear-work EREW PRAM algorithms for spanning
forests, bridges, and biconnected components.

For biconnectivity, Tarjan and Vishkin~\cite{TV85} gave a deterministic
\(O(\log n)\)-time CRCW PRAM algorithm using \(O(m+n)\) processors;
their framework also finds all bridges. Subsequent work has studied
the practical performance of these decompositions~\cite{EV12}.
Weighted spanning forests have a separate line of work: Cole, Klein,
and Tarjan~\cite{CKT96} obtained logarithmic time and linear work on
the CRCW PRAM, and Pettie and Ramachandran~\cite{PR02} achieved these
bounds on the EREW PRAM. Although these weighted results offer helpful
benchmarks, they do not improve the time bound for graphs with a small
diameter, which is typical in most real-world applications.

The massively parallel computation (MPC) model is generally regarded
as more powerful than the PRAM~\cite{KSV10,LTZ20}. Andoni et al.~\cite{ASSWZ18} gave an
\(O(\log d\,\log\log n)\)-round connectivity algorithm in
low-space MPC for sparse graphs.
Assadi, Sun, and Weinstein~\cite{ASW19} gave an
\(O(\log(1/\lambda)+\log\log n)\)-round MPC connectivity algorithm
for sparse graphs, where \(\lambda\) is the minimum component
spectral gap.
Behnezhad et al.~\cite{BDELM19} subsequently improved the diameter-based bound to
\(O(\log d+\log\log n)\) rounds, and Coy and
Czumaj~\cite{CC22} matched this bound deterministically.
For forest inputs, Balliu et al.~\cite{BLMOU23} obtained
deterministic \(O(\log d)\)-round
connectivity and forest rooting with linear total memory.

Liu, Tarjan, and Zhong (LTZ)~\cite{LTZ20} obtained
\(O(\log d+\log\log n)\)-time connectivity on the ARBITRARY CRCW
PRAM using \(O(m+n)\) processors, with good probability.\footnote{\emph{With good probability} means with probability at least \(1-1/\poly(((m+n)\log n)/n)\), which is at least \(1 - 1/\poly(\log n)\). In particular, whenever \(m=n^{1+\delta}\) for any constant \(\delta>0\), the success probability is \(1-n^{-\Omega(1)}\), which is high probability.}
Farhadi, Liu, and Shi~\cite{FLS24} subsequently obtained
connectivity in \(O(\log(1/\lambda)+\log\log n)\) time and optimal
\(O(m+n)\) work with high probability, where \(\lambda\) is the
minimum component spectral gap.

%For spanning forests, the corresponding bounds remain multiplicative.
As for spanning forest, in low-space MPC with linear total memory, Andoni et al.~\cite{ASSWZ18}
and Coy and Czumaj~\cite{CC22} give
\(O(\log d\,\log\log n)\)-round algorithms on sparse graphs;
the latter is deterministic. LTZ~\cite{LTZ20} obtain the same time
bound with good probability on the ARBITRARY CRCW PRAM.
Fast connectivity algorithms create virtual edges that may represent
long paths in the input graph, but does not retain the information
required to recover their original edges. A spanning-forest algorithm
must preserve this information and recover a set of original edges
that spans every component without creating cycles. The central
challenge is to produce this stronger output within the
\(O(\log d+\log\log n)\) bound.

For other graph problems closely related to computing connected
components, Andoni, Stein, and Zhong~\cite{ASZ19} gave low-space MPC
algorithms for finding bridges and computing biconnected blocks.
On sparse graphs with linear total memory, their algorithms take
\(O(\log d\,\log\log n)\) rounds for bridges and
\(O(\log d\,(\log\log n)^2+\log d'\,\log\log n)\) rounds for
biconnectivity, where \(d'\) is the bi-diameter.

Our results obtain improved randomized PRAM algorithms for all these fundamental graph problems. In particular, we compute a spanning forest and find all bridges in $O(\log d+\log\log n)$ time, and compute all biconnected blocks in $O(\log d+\log d'+\log\log n)$ time, where $d$ is the maximum component diameter and $d'$ is the bi-diameter. All algorithms use $O(m+n)$ processors and succeed with good probability. These results improve the best previously known time bounds while retaining linear processor.

\subsection{LTZ Review}
\label{sec:ltz-review}
The LTZ algorithms~\cite{LTZ20} use hashing for connectivity and
spanning forests on the ARBITRARY CRCW PRAM.
In each phase, the algorithm repeatedly squares a budget assigned to
every current vertex, and because a larger budget can absorb more nearby
vertices, the number of current vertices drops quickly. It expands each
neighborhood by doubling the distance that can be inspected, chooses a
leader, and contracts the resulting links. For a spanning forest, it
follows the saved original-edge links backward after all contractions
finish.

For connected components, LTZ use a faster routine that does not retain
original-edge witnesses. Classical PRAM connectivity bounds depend on
\(n\), whereas this routine makes progress in terms of the current
diameter. Its key advantage is level-dependent processor allocation: each
representative receives only the processors required by its current
level, rather than the same allocation as every vertex. This keeps more
of the linear processor pool active on useful expansion work and reduces
idle capacity. Each representative stores a level, a budget, and a
hashed neighbor table. Star contractions reduce the number of
representatives; \textsc{Maxlink} redirects them toward higher-level
representatives, \textsc{Shortcut} skips intermediate parents, and
\textsc{Alter} updates edge endpoints after a parent changes. Two-hop
expansion doubles the explored radius in each phase. These techniques yield
\(O(\log d+\log\log n)\) time for connected components. Retaining the
original edges for a spanning forest adds the \(\log\log n\) factor in
the sparse regime, giving \(O(\log d\,\log\log n)\) time.

\subsection{Main Contribution}
\label{sec:contribution}
Our central contribution is an original-edge spanning-forest algorithm
whose diameter dependence is additive rather than multiplicative: it
runs in \(O(\log d+\log\log n)\) time with a linear number of
processors and returns a bounded-depth rooted forest. Using this
forest, we obtain the same additive bound for bridge identification and
an \(O(\log d+\log d'+\log\log n)\) bound for biconnected blocks,
without computing a DFS Sequence. We also establish lower bounds for
DFS Sequence and weighted spanning-forest problems.

Let
\(G=(V,E)\) be an undirected graph with \(n\) vertices, \(m\) edges,
and maximum component diameter \(d\). Let \(d'\) denote its
bi-diameter: the maximum, over pairs of vertices on a common simple
cycle, of the length of a shortest simple cycle containing the pair.
Set \(d'=1\) when the graph has no cycle.
Every logarithmic factor is at least one. The input has no self-loops;
loops introduced by our implementations are bookkeeping records only.
Parallel edges are allowed and retain distinct input identifiers.
As usual in the word model, \(m=n^{O(1)}\).

A parallel random access machine (PRAM) consists of processors that
operate synchronously and communicate through shared memory.
We use ARBITRARY CRCW PRAM model: processors may read or write the same memory cell
simultaneously, and an arbitrary writer succeeds in each write conflict.
Each of the \(O(m+n)\) processors has \(O(1)\) private words of
\(O(\log n)\) bits and performs \(O(1)\) word operations per step.
Shared memory need not be linear in the input size. 

% 介绍一下模型，bridge indicator改成finding bridges
\begin{theorem}[Spanning Forest]
\label{thm:main-sf}
There is a randomized ARBITRARY CRCW PRAM algorithm that, with good probability, outputs an
original-edge rooted spanning forest in
\(O(\log d+\log\log n)\) time using \(O(m+n)\) processors. The maximum rooted depth of the
output is \(2^{O(\log d+\log\log n)}\).
\end{theorem}

\begin{theorem}[Finding bridges]
\label{thm:main-bridge}
There is a randomized ARBITRARY CRCW PRAM algorithm that, with good probability, outputs an edge-indexed Boolean array identifying all bridges in
\(O(\log d+\log\log n)\) time using \(O(m+n)\) processors.
\end{theorem}

\begin{theorem}[Biconnectivity]
\label{thm:main-block}
There is a randomized ARBITRARY CRCW PRAM algorithm that, with good probability, outputs an edge-indexed labeling of the biconnected blocks in
\(O(\log d+\log d'+\log\log n)\) time using \(O(m+n)\) processors.
\end{theorem}

For spanning forests, the best previous diameter-sensitive bound with the
same linear processor restriction is
\(O(\log d\,\log\log n)\) in the sparse regime~\cite{LTZ20}.
Our result makes the \(\log\log n\) factor additive and matches
their density-independent connected-components bound for the stronger
output of an original-edge rooted forest. On dense inputs with
\(m=n^{1+\Omega(1)}\), LTZ already obtain \(O(\log d)\) time;
our improvement concerns the sparse, density-independent bound.
For each nonroot vertex, the output
specifies its parent and the corresponding input-edge identifier;
the maximum rooted depth is at most
\(2^{O(\log d+\log\log n)}\).
The previous low-space MPC spanning-forest algorithms have the same
multiplicative time dependence, including the deterministic result
reported in~\cite{CC22}. The comparison is to
spanning-forest construction, rather than to the faster algorithms
that return only component labels.

\begin{table}[htbp]
\centering
\scriptsize
\setlength{\tabcolsep}{3pt}
\renewcommand{\arraystretch}{1.3}
\resizebox{\linewidth}{!}{%
\begin{tabular}{|c|c|c|c|}
\hline
\textbf{Problem} & \textbf{PRAM} & \textbf{MPC} & \textbf{This paper}\\
\hline
Spanning forest & \mbox{\(O(\log d\,\log\log n)\)~\cite{LTZ20}} &
\mbox{\(O(\log d\,\log\log n)\)~\cite{ASSWZ18}} &
{\boldmath\(O(\log d+\log\log n)\)}\\
\hline
Finding bridges & \mbox{\(O(\log n)\)~\cite{TV85}} &
\mbox{\(O(\log d\,\log\log n)\)~\cite{ASZ19}} &
{\boldmath\(O(\log d+\log\log n)\)}\\
\hline
Biconnectivity & \mbox{\(O(\log n)\)~\cite{TV85}} &
\mbox{\(O(\log d\,\log\log n+\log d')\)~\cite{ASZ19,BDELM19}} &
{\boldmath\(O(\log d+\log d'+\log\log n)\)}\\
\hline
\end{tabular}
}
\caption{Comparison of time bounds.}
\label{tab:main-comparison}
\end{table}

Here the table is for the sparse regime \(m=O(n)\). PRAM bounds use
\(O(m+n)\) processors; low-space MPC bounds use \(O(n^\delta)\)
local memory, for any fixed \(0<\delta<1\), and \(O(m+n)\) total
memory. The biconnectivity MPC entry combines the reduction~\cite{ASZ19}
with the later connectivity algorithm~\cite{BDELM19}. Bounds depending on
diameter can also be capped by the classical \(O(\log n)\) bounds.

For bridges and biconnected blocks, the classical PRAM bounds are
\(O(\log n)\)~\cite{TV85,HZ01}. Our bounds instead depend on the
input geometry. In the sparse, linear-total-space MPC regime,
the MPC results~\cite{ASZ19} give
\(O(\log d\,\log\log n)\) rounds for bridges and
\(O(\log d\,(\log\log n)^2+\log d'\,\log\log n)\) rounds
for blocks. Combining this biconnectivity reduction with the later
connectivity algorithm~\cite{BDELM19} gives the
stronger comparison bound
\(O(\log d\,\log\log n+\log d')\).
Our result replaces the remaining multiplicative
\(\log\log n\) dependence by an additive term, while operating on
the more restrictive PRAM. For example, when
\(d=\log^{\Theta(1)}n\), our spanning-forest and bridge bounds
are \(O(\log\log n)\), compared with the previous
\(O((\log\log n)^2)\) sparse diameter-sensitive bounds.
For biconnectivity, the \(d'\) parameter remains relevant even when
\(d\) is constant: an MPC lower bound of \(\Omega(\log d')\) is known
under the one-cycle versus two-cycles conjecture~\cite{ASZ19}.

Our lower-bound contributions are stated separately.

\begin{theorem}[Lower Bounds]
\label{thm:main-lower-bounds}
On an ARBITRARY CRCW PRAM with a linear number of processors, every deterministic algorithm for any of the
following problems requires
\[
  \Omega\!\left(\frac{\log n}{\log\log n}\right)
\]
worst-case time:
\begin{enumerate}[label=(\roman*),leftmargin=1.6em,itemsep=3pt,parsep=0pt]
\item outputting a DFS Sequence array from parent-array input(even
when the input tree has depth two and the child order is unrestricted);
\item outputting an exact minimum spanning forest(even on connected
diameter-two graphs);
\item outputting a fixed-factor approximation to a minimum spanning
forest or to a bottleneck minimum spanning forest(even on connected
diameter-two graphs with two distinct edge weights).
\item outputting an open ear decomposition.
\end{enumerate}
\end{theorem}

%KS 年份
Part~(i) follows from Corollary~\ref{cor:dfs}; Parts~(ii)--(iii) follow from Theorem~\ref{thm:mst-lower}; Part~(iv) follows from Theorem~\ref{thm:parity-to-oed}. The deterministic statements use the
classical PARITY lower bound of Beame and H{\aa}stad~\cite{BH89}.
Randomized consequences require a PARITY lower bound with matching
model and error assumptions~\cite{KSParity}. These restrictions on
traversal and weighted outputs distinguish them from the original-edge
spanning forests and indexed decomposition labels computed here.

\subsection{Technique Overview}
\label{sec:technique-overview}
Three ideas support our algorithms: shared witnesses for virtual
edges, rank-preserving recovery of original-edge parents, and shared
path marking in place of subtree aggregation.

\paragraph{Immutable witnesses for virtual edges.}
Virtual edges preserve connectivity but may hide long routes through
the original graph. Copying these routes whenever an edge is reused
would lose the resource bounds. We instead give each virtual arc a
reference to an immutable, constant-size record identifying either
an original arc or two earlier witnesses whose walks concatenate.
Many later records can share the same earlier witness, so composing
or copying a route requires only constant additional work.
Each vertex retains the witness of its current parent. At termination,
these references describe the final parent forest, and recovery
follows the shared records backward until original arcs are reached.

\paragraph{Rank-preserving recovery and depth control.}
Witness walks may meet at the same intermediate vertex, so expanding
them independently can create conflicting parents or cycles.
We coordinate their choices through an integer rank that strictly
increases along every nonroot parent arc. A grouped maximum selects
each new rank together with its parent witness. The selection rule
preserves the roots and components while maintaining strict rank
increase, so each reverse layer directly returns a rooted forest.
The same invariant controls depth: recovery starts from a certified
star forest, and the largest rank grows by at most a constant factor
per layer. Rooting each preparation tree at its attachment vertex
then extends this bound to the original graph.
This depth guarantee keeps the subsequent bridge and block
processing within the additive time bound.

\paragraph{From subtree aggregation to shared path marking.}
Classical bridge and block algorithms aggregate edge information over
subtrees in traversal order~\cite{TV85,ASZ19}.
We reformulate the required subtree conditions as path-marking
problems, avoiding a DFS Sequence. Expanding each path separately
would create too many tasks when paths overlap. We instead register
representative marks in a shared hierarchy of tree segments, where
all requests for the same segment merge into one mark. The marks
are then expanded level by level, combining overlaps before
continuing to shorter segments and finally individual tree edges.
Storage is allocated to shared segments rather than their separate
occurrences in input paths, and each expansion level reuses the
same vertex processors. This yields the bridge tests and auxiliary
edge conditions deterministically in \(O(\log h)\) time with
\(O(m+n)\) processors for a rooted forest of depth \(h\).

Rank aggregation uses classical prefix dictionaries and
hashing~\cite{Willard83,CW79,DHKP97,GM94}. Conditional on the
forward resource and round bounds, all inverse layers compute their
grouped maxima in constant time per layer with high probability,
using the same linear processor pool. The aggregation primitives
are standard; the recovery analysis establishes the simultaneous
time guarantee needed here.

\section{Spanning forests by witnessed expansion}
\label{sec:additive-sf}
We improve the connectivity framework of Liu, Tarjan, and
Zhong~\cite{LTZ20} to compute spanning forests of bounded depth.
The forward construction uses virtual edges to reduce distances.
Its loop stops when local tests can join the remaining roots by
stars. A terminal sequence of pointer shortcuts then gives a witnessed
star forest on all retained vertices. Recovery starts from this final parent array
and reverses its witness layers.
This gives an initial depth of one and a simple rank bound for the
recovered forest. Only the current parent witness of each vertex is
needed, together with the immutable witness history.

\subsection{Shared state and the expansion schedule}
\label{sec:hash-schedule}
Pad the input with one loop per vertex and put \(M=m+n\).
The shared array \(p[v]\) stores temporary parents; \(\Par[v]\) is
reserved for original-edge output parents.
A root satisfies \(p[v]=v\), and the parent forest is flat when
\(p[p[v]]=p[v]\) for every vertex.
The current undirected graph is represented by both orientations
of every retained arc. It includes altered input arcs and all
retained expansion arcs. Write \(N(v)\) for the current neighbors
of \(v\), including \(v\).

\paragraph{Workspace preparation.}
Fix a sufficiently large constant \(c\).
Preparation and allocation formulas use the integer ceiling of
$\log_2 n$ for capacities and loop bounds; this affects only constants.
If \(M/n\le\log^c n\), run
\(R_0=\lceil c\log_{8/7}\log n\rceil\) star-contraction phases.
In each phase, every current root independently becomes a center
with probability \(1/2\). A noncenter with a center neighbor selects
one incident input arc by an arbitrary write and links to that center.
Record the selected original arc before moving its endpoints.
One pointer-shortcut makes the temporary parents flat again;
move every current arc to its endpoint parents and save the resulting
owner row. These operations take constant time per phase.
If \(M/n>\log^c n\), take \(R_0=0\).
Let \(F_0\) be the selected preparation forest. It is a forest by
induction: each phase joins disjoint existing trees along a star,
using one original edge per noncenter component.
Components with no remaining nonloop edge are already spanned by
\(F_0\) and take no further part in the expansion schedule.

On sparse inputs, approximately compact the remaining representatives
once into a padded range of capacity
\(q=\lceil2M/\log^c n\rceil\).
On dense inputs, retain their input indices and set \(q=n\).
Store \(g(v)\), the unique slot of each retained representative \(v\).
The slot map preserves the original graph labels. Keep the compacted
slot array as well: slot $g(v)$ contains the identifier of $v$, and
holes have a validity bit. Both maps remain available after a vertex
ceases to be a root. No exact output count is required.
The primitive takes \(O(\log^* n)\) time with \(O(n)\) processors;
on a length-\(q\) range it takes constant time with \(O(q\log q)\)
processors, using the approximate-compaction primitive invoked
in~\cite{LTZ20}. The initial compaction is therefore paid once, and
subsequent allocations fit within the linear pool.
In particular, \(q\log^2 n=O(M)\).
The selected-arc array and owner rows are the records used by
Algorithm~\ref{alg:restore}.

The remaining representatives form the base graph \(G_0\).
Initialize their parents to themselves and levels $\lambda[v]$ to
one. The nominal initial budget is
$\max\{M/n,\log^c n\}/\log^2 n$. For a finite-word implementation,
round $\log n$ to its integer ceiling and round this budget upward
to a power of four. Write the resulting budget as $b_1=2^{2\beta_1}$,
and precompute
\[
 \beta_{j+1}=\lceil101\beta_j/100\rceil,
 \qquad b_j=2^{2\beta_j}.
\]
Then $b_j^{1.01}\le b_{j+1}<4b_j^{1.01}$, and a root's neighbor
table has exactly $2^{\beta_j}=\sqrt{b_j}$ slots. A root of level
$j$ requests a block of size $b_j$. Its sampling probability is
$2^{-\lceil3\beta_j/25\rceil}$, which lies between
$b_j^{-0.06}/2$ and $b_j^{-0.06}$. Sampling tests whether a fixed
number of bits in an independent random word are zero.

Budgets and sampling exponents are stored in shared level-indexed
tables. Precompute them until the next budget exceeds the fixed
processor pool, then store an overflow sentinel. There are
$O(\log\log n)$ entries because $M$ is polynomial in $n$ and
$\beta_j$ grows geometrically. Each next entry uses constant-time
integer arithmetic. The initial exponent is found in constant time
by $O(\log n)$ processors testing consecutive powers of four against
the two inequalities defining the initial budget; multiplication clears
the denominators. Integer logarithms are found similarly by testing
consecutive powers of two. All tested quantities have $O(\log n)$
bits for fixed $c$. Thus table preparation takes $O(\log\log n)$
time, including the dense-input case, and no real exponentiation is
performed within an expansion round.

Write $b(v)$ for the size of the block already allocated to $v$.
Initially every retained root receives a block of size $b_1$.
Only roots increase levels; nonroot levels remain fixed.
The budget stays unchanged throughout a round until its final
allocation. In particular, sampling changes a root's level but not
the table size used in that round. A sampled root that subsequently
becomes a nonroot keeps its old block and frozen level; it receives
no new block. All vertex-indexed operations skip holes in the range
of $g$. The constant-factor rounding is accounted for in the forward
analysis below.

\paragraph{Basic operations.}
Algorithm~\ref{alg:basic-ops} specifies the three parent and edge
operations used in the expansion loop.
Each simultaneous update reads the old shared state and writes to
a separate output array. Before publishing a changed parent or arc,
we construct its witness from the same old state, as described in
\S\ref{sec:sf-witnesses}. The new fields and their witness reference
are then published together.

\noindent\begin{minipage}{\linewidth}
\begin{framed}
\small
\setlength{\parindent}{0pt}
\algcaption{Basic Graph Operations}
\label{alg:basic-ops}\par
\smallskip
\textbf{Input:} Shared parents \(p\), levels \(\lambda\), and
current arcs.\par
\textbf{Output:} Updated parents or arc endpoints.\par
\smallskip
\begin{enumerate}[label=\arabic*.,ref=\arabic*,leftmargin=1.5em,topsep=3pt,itemsep=3pt,parsep=0pt]
\item \label{step:basic-maxlink}
\textsc{Maxlink}: Repeat twice.
\begin{enumerate}[label=\alph*.,ref=\arabic{enumi}.\alph*,leftmargin=1.4em,topsep=2pt,itemsep=2pt,parsep=0pt]
\item \label{step:maxlink-select}
For each vertex \(v\), choose
\(w\in\argmax_{u\in N(v)}\lambda[p[u]]\).
Set \(p_{\rm new}[v]\gets p[w]\) if
\(\lambda[p[w]]>\lambda[v]\), and
\(p_{\rm new}[v]\gets p[v]\) otherwise.
\item \label{step:maxlink-publish}
Replace \(p\) by \(p_{\rm new}\).
\end{enumerate}
\item \label{step:basic-shortcut}
\textsc{Shortcut}: For each vertex \(v\), set
\(p_{\rm new}[v]\gets p[p[v]]\). Replace \(p\) by \(p_{\rm new}\).
\item \label{step:basic-alter}
\textsc{Alter}: Replace each current arc \((u,v)\) by
\((p[u],p[v])\), using a separate output array.
\end{enumerate}
\end{framed}
\end{minipage}\par\smallskip

\textsc{Maxlink} selects a neighbor's parent of maximum level.
\textsc{Shortcut} replaces each parent pointer by a pointer to its grandparent.
\textsc{Alter} moves arc endpoints to their parents and retains
the arc identifiers.
The expansion round combines these operations with neighbor hashing
and the termination test.
The test uses shared root-indexed arrays:
\(\mathsf{center}[u]\) marks a candidate center,
\(\mathsf{pick}[u]\) stores its selected arc to a center, and
\(z[u]\) stores its chosen center identifier.
A failed termination requirement returns \textsc{Continue} before
any terminal parent update.

\noindent\begin{minipage}{\linewidth}
\begin{framed}
\small
\setlength{\parindent}{0pt}
\algcaption{Expansion Round}
\label{alg:expand}\par
\smallskip
\textbf{Input:} Shared parents, levels, current arcs, and processor
blocks, with parent and arc witnesses.\par
\textbf{Output:} Updated state and status \textsc{Continue} or
\textsc{Finished}, or \textsc{Failure} if a resource check fails.\par
\smallskip
\begin{enumerate}[label=\arabic*.,ref=\arabic*,leftmargin=1.5em,topsep=3pt,itemsep=3pt,parsep=0pt]
\item \label{step:expand-link}
Execute \textsc{Maxlink} and \textsc{Alter}; clear dormant flags.

\item \label{step:expand-sample}
Each root \(v\) independently increases its level by one with
the precomputed probability for its current budget. Record the choice.
Choose a fresh pairwise independent hash function.

\item \label{step:expand-hash}
For each root \(v\), hash roots \(w\in N(v)\) with
\(b(w)=b(v)\) into a fresh table \(H_{\rm old}(v)\).
Mark a root dormant if its table has a collision or lists a root
whose table has a collision.

\item \label{step:expand-neighbors}
For each root \(v\), copy \(H_{\rm old}(v)\) into its fresh output
table and also hash every \(u\in H_{\rm old}(w)\), \(w\in H_{\rm old}(v)\), into
that output table, using the same hash function.
Mark \(v\) dormant if this output table has a collision.
Execute \textsc{Maxlink}, \textsc{Shortcut}, and \textsc{Alter}.

\item \label{step:expand-allocate}
Increase each remaining dormant root's level unless it already
increased this round. For each remaining root $v$, allocate a fresh
block of size $b_{\lambda[v]}$ and update its current budget.

\item
Test termination, ignoring self-loops.
\begin{enumerate}[label=\alph*.,ref=\arabic{enumi}.\alph*,leftmargin=1.4em,topsep=2pt,itemsep=2pt,parsep=0pt]
\item \label{step:close-flat}
Require root endpoints for every nonloop current arc.
Set \(\mathsf{center}[u]\gets1\) and
\(\mathsf{pick}[u]\gets\nil\) for every root \(u\).

\item \label{step:close-centers}
Each arc \((u,v)\) with \(v<u\) writes zero to
\(\mathsf{center}[u]\).
Each remaining center \(u\) sets \(z[u]\gets u\).

\item \label{step:close-proposals}
Each arc from a noncenter \(u\) to a center \(v\) proposes its
identifier to \(\mathsf{pick}[u]\).
Require a proposal for every noncenter.
Each noncenter sets \(z[u]\gets v\) using its selected arc.

\item \label{step:close-check}
Require \(z[u]=z[v]\) on every current arc.
\end{enumerate}

\item \label{step:close-finish}
For each noncenter root \(u\), set
\(p[u]\gets z[u]\) and copy the witness of \(\mathsf{pick}[u]\).
Keep other parents unchanged.

\item \label{step:close-flatten}
While some retained vertex $v$ has $p[p[v]]\ne p[v]$, execute
\textsc{Shortcut} and record its witnesses.
Return \textsc{Finished}.
\end{enumerate}
\end{framed}
\end{minipage}\par\smallskip

\paragraph{Executing a round.}
At the start of each \textsc{Maxlink} iteration, clear the shared
candidate cells and rejection flags. Each arc processor proposes its
record identifier to the cell indexed by its tail and the level of its
head's current parent. Each retained vertex $v$ also proposes its
identity witness $v\to v$ to the cell for level $\lambda[p[v]]$.
This explicit self-candidate implements $v\in N(v)$ even when its
original loop has moved elsewhere under \textsc{Alter}.
It is a temporary proposal, not a new retained current arc.
All candidates in one cell have the same level. One processor per pair
of level indices rejects smaller occupied levels; the largest publishes
its winning record identifier. Its head specifies $w$, and its witness
provides $v\to w$. All these reads use the same parent snapshot.
There are $O(\log n)$ levels, so clearing, selecting, and publishing use
$O(q\log^2 n)=O(M)$ processors and constant time per iteration.
No combining maximum write is used.

For expansion, processor \((v,a,b)\) reads slot \(a\) of
\(H_{\rm old}(v)\), obtains \(w\), and reads slot \(b\) of
\(H_{\rm old}(w)\).
Equal budgets ensure \(\sqrt{b(v)}\sqrt{b(w)}=b(v)\), so these are exactly the
processors in \(v\)'s block. A collision is detected by writing one
record identifier, reading the winner, and comparing vertex keys.
A losing distinct key writes one to that table's collision flag.
The next pass propagates the table-collision flags once; it does not
recursively propagate newly set dormant flags.
Repeated copies of the same key are harmless. Copied old entries
and newly generated two-hop candidates participate in the same
write-and-readback collision test. In particular, a new key overwriting
an old key must mark a collision even if no two new keys collide.
There are $O(\sqrt{b(v)}+b(v))=O(b(v))$ such tasks per block.
No two-hop tasks are counted or compacted.
Each root also hashes its identity reference in the first stage,
so the implicit self-neighbor is represented explicitly.

For allocation, reserve a separate processor zone for every pair
consisting of a round number and a level. Each zone contains equal-size
blocks for that level. Approximately compact its roots within the
reduced vertex range; a root's resulting position directly identifies
its block, without a prefix sum. The output capacity is the known
capacity of the zone; the call needs only this upper bound, not an exact
count of its roots. Old blocks remain allocated, including
the tables whose entries represent retained arcs. Each block holds a
constant number of tables and at most a constant multiple of its size
in arc cells. Every retained arc has one processor, up to a constant
factor for its two orientations. These processors update its endpoints
in \textsc{Alter} and scan it in \textsc{Maxlink} and the termination
test. An immutable witness record is not a new current arc unless its
identifier wins a retained table slot.

The implementation fixes a linear processor pool and zones for at most
$A\log n$ rounds and levels, for a sufficiently large constant $A$.
Allocation sizes are capped at the pool size plus one; an overlarge
request is a failure signal, not an unbounded integer or an allocation.
Compaction is run with the corresponding fixed time and output-capacity
limits. Every root checks that its assigned slot is in range and
unique by writing and reading back its identifier; failures set one
shared flag. An out-of-range allocation or an exhausted round limit
returns \textsc{Failure} before any out-of-range access. Thus the
processor bound is unconditional. The following lemma bounds the
probability of such failures and the time on successful executions.

\begin{lemma}[Forward connectivity invariants]
\label{lem:forward-invariants}
Before the terminal parent update, the temporary parents form a forest.
The graph consisting of current arcs and parent arcs has the same
components as \(G_0\). Every vertex incident to a nonloop current
arc is connected to its parent by a path of current arcs.
\end{lemma}
\begin{proof}
Levels strictly increase along every nonroot parent arc:
\textsc{Maxlink} selects a larger level, only roots increase levels,
and \textsc{Shortcut} follows two increasing arcs.
Thus the parent graph is a forest.
The other claims hold initially, when all parents are self-pointers.

Consider a \textsc{Maxlink} update at an incident vertex \(v\),
using neighbor \(w\). Before the update, current paths connect
\(v\) to its old parent and \(w\) to \(p[w]\).
Hence both its old and new parent arcs lie within the same current
component. Changing these arcs preserves both claims.
An isolated vertex does not change its parent.
\textsc{Shortcut} preserves the parent-tree components and extends
the asserted current path through the old parent's current path.

In \textsc{Alter}, replacing \((u,v)\) by \((p[u],p[v])\)
preserves connectivity when parent arcs are included.
If \(x=p[u]\) is incident to a new nonloop arc, applying the same
endpoint map to the old current path from \(u\) to \(p[u]\)
gives a current walk from \(x\) to \(p[x]\).
Expansion only adds connections already supplied by current paths.
These observations prove the invariants by induction.
\end{proof}

\begin{lemma}[Forward schedule and its resource bound]
\label{lem:hash-schedule}
On a good-probability event $\mathcal E$, workspace preparation takes
$O(\log\log n)$ time and leaves $q=O(M/\log^c n)$ representative
slots. The forward operations and termination test in each of the
first $A\log n$ rounds take constant time, excluding terminal flattening.
All blocks allocated across these rounds together use $O(M)$ processors,
and all retained current arcs occupy $O(M)$ slots.
Within $O(\log d+\log\log n)$ rounds, every nontrivial component of
the current graph has diameter at most one. If the algorithm has not
already terminated, the next round ends with root endpoints on all
nonloop arcs and clique components in the current root graph.
\end{lemma}
\begin{proof}
We apply the processor-allocation and diameter-reduction analysis of
the faster connectivity schedule in the full version of~\cite{LTZ20}.
The correspondence
is as follows. Our preparation uses the same independent center votes,
star links, one shortcut, and endpoint alteration. After preparation,
our $\lambda[v]$, $b(v)$, and $H(v)$ are respectively its levels,
current budgets, and neighbor tables. The two \textsc{Maxlink} calls,
the shortcut, the two alterations, and the dormancy rules in
Algorithm~\ref{alg:expand} are unchanged. Budgets and probabilities
use the finite-word rounding specified above; we check its effect on
the quantitative bounds below. In particular, both hashing
stages read their old tables, and the second retains the old keys.
The current graph includes all previously retained table arcs, exactly
as in that schedule. Preparation nonroots have no incident current
nonloop arcs and cannot influence the subsequent expansion; we omit
them from $G_0$ and retain their history for final recovery.
Contracting the preparation trees does not increase component diameter.

Fix the random choices and an arbitrary resolution of concurrent writes.
The witness fields do not enter any vote, hash key, comparison, or
parent choice. A collision is still resolved by its winning vertex key;
if several records have that key, any one of their valid witnesses
may be retained. Failed termination tests change no forward field.
Consequently, removing witness annotations gives exactly the same
execution of this rounded forward schedule, up to the first successful
certificate. Its correspondence with the cited analysis rests on the
unchanged transitions and the inequalities checked below, rather than
on equality of the unrounded numerical probabilities. The omitted preparation nonroots may still be shortcutted in that
comparison execution, but have no current nonloop arcs and cannot affect
the retained state. Only its diameter-one and allocation bounds are used below.
We do not require its stopping test or a simultaneous flattening event.
No new assumption about independent states in successive rounds is made.

For completeness, we spell out how its resource bounds apply here.
In preparation, an ongoing representative has a neighbor and retires
with probability at least $1/4$ per phase. Its survival probability
after $R_0$ phases is at most $(3/4)^{R_0}$. Markov's inequality gives
at most $(7/8)^{R_0}n\le M/\log^c n$ ongoing representatives except
with probability $(6/7)^{R_0}$. In the sparse regime this is inverse
polynomial in $(M\log n)/n$. Approximate compaction gives the padded
range $q$; the dense regime uses $q=n$. Recording original arcs and
owner rows adds $O(1)$ work per edge or vertex in each phase, so does
not change its time or processor bound.

Here is the quantitative effect of rounding. For a root with current
budget $b$, if some neighbor has at least $b^{0.1}$ neighbor roots of
that budget, the probability that none is sampled is at most
$\exp(-b^{0.04}/2)$. Otherwise the two hashing stages and dormancy
propagation give collision bounds
$O(b^{-0.3}+b^{-0.2}+b^{-0.1})$, as in the allocation argument
of~\cite{LTZ20}. Including the root's own sampling, its probability of
receiving a larger block in this round is at most $b^{-0.05}$ for
sufficiently large $c$. Constant factors in the hash collision bound
and the rounded sampling probability are absorbed by this exponent.

We make the counting argument explicit to avoid conditioning on a
population selected by future random choices. Let
$Z=C M/\log^2 n$ be a logical load limit, for a sufficiently large
constant $C$, and let $a_j=\lfloor Z/b_j\rfloor$.
Preparation ensures $n_1b_1\le Z$ on its successful event.
For each level $j$, consider only the first $a_j$ representatives ever
allocated budget $b_j$, ordering simultaneous arrivals by identifier
for analysis only. Conditional on any such representative's arrival
history, its probability of receiving the next budget within the
remaining $A\log n$ rounds is at most
$A\log n\,b_j^{-0.05}\le b_j^{-0.04}$.
Thus the expected number $Y_j$ of promotions among these first arrivals
is at most $a_jb_j^{-0.04}$, without any independence assumption.
If level $j$ never exceeds $a_j$ arrivals but level $j+1$ exceeds
$a_{j+1}$, then $Y_j\ge a_{j+1}+1$. Markov's inequality gives
\[
 \Pr[Y_j\ge a_{j+1}+1]
 \le \frac{a_jb_j^{-0.04}}{a_{j+1}+1}
 \le \frac{b_{j+1}}{b_j}\,b_j^{-0.04}
 \le 4b_j^{-0.03}\le b_j^{-0.02}.
\]
The argument also covers $a_{j+1}=0$: a single promotion then causes
failure. Taking a union bound over the first level whose load limit
is exceeded proves that every attained level has total load at most
$Z$, except with probability $O(\log n)b_1^{-0.02}$.
For sufficiently large $c$, this is inverse polynomial in
$(M\log n)/n$. These counts and their arrival order are used only
in the proof and are never computed by the PRAM.

Reserve a physical zone of size $2Z$ for each round--level pair.
Each compaction call is given an output range of capacity $2a_j$ in
that zone. At any round the number of roots being assigned is at most the total number of
arrivals at that level, hence at most $a_j$ on the allocation event.
At most $O(\log^2 n)$ zones are used, and the factor-two slack for
approximate compaction is absorbed in their constants. Thus the sum
of all old and new allocated blocks is $O(M)$. Compacting the roots
at all levels simultaneously takes constant time with
$O(q\log q\log n)=O(M)$ processors. The maximum-level selection
uses $O(q\log^2 n)=O(M)$ processors, and the two-hop tasks use their
already assigned blocks. These are a constant number of linear pools,
which together still have size $O(M)$.

Each retained arc occupies a table slot in one of these allocated
blocks, or an original input slot. Therefore there are $O(M)$ current
arc slots over the entire forward execution, even though many
immutable witness versions may describe them. Scanning all current
arcs in every round costs constant time with $O(M)$ processors.
The immutable records instead take $O(MS)$ shared words for $S$
witness layers; they are not all scanned in a forward round.

On the allocation event, $b_j\ge b_1^{1.01^{j-1}}$ and no
allocated budget exceeds $O(M)$, so every attained level is
$O(\log\log n)$. A nonroot may have a level one higher than its
last allocated budget level, which changes this bound only by one.
The path-reduction argument of~\cite{LTZ20} uses this level bound,
strict level increase on parent arcs, and the following expansion
property: a root that stays a root without increasing its level
retains all relevant two-hop neighbors. This property is unchanged
by rounding. Such a root and the neighbor parents used by expansion
have equal current budgets; a missing key would trigger dormancy and
a level increase. The same path argument therefore gives diameter
one in $O(\log d+\log\log n)$ rounds. We next justify the extra
root-endpoint property directly, instead of strengthening that imported
result to a simultaneous flatness statement.
Consider a nontrivial current clique $Q$ at the start of the next round.
By Lemma~\ref{lem:forward-invariants}, the parent of every vertex of $Q$
also lies in $Q$. A vertex of maximum level in $Q$ must therefore be a
root, since a nonroot has a parent of strictly larger level.
Every vertex of $Q$ sees such a root in its neighborhood. Thus the first
\textsc{Maxlink} sends every lower-level vertex of $Q$ directly to a
maximum-level root, while those roots remain fixed. \textsc{Alter}
maps the clique onto a clique on these roots.

Subsequent neighbor expansion adds only edges within a current clique.
After sampling, the second \textsc{Maxlink} again sends every incident
lower-level vertex to a maximum-level root of that clique. Shortcutting
does not change these root destinations; the last \textsc{Alter}
therefore leaves a clique with root endpoints. Dormancy updates change
levels, not parents or endpoints. Loops and isolated vertices do not
affect this argument. It holds in all components in the same round.
The cited allocation, compaction, and diameter events have good
probability, and on their intersection no resource limit is hit.
\end{proof}

\paragraph{The termination test.}
The final part of Algorithm~\ref{alg:expand} checks whether each
root component can be joined by a star.
The center flag means that no current neighbor has a smaller
identifier. All temporary arrays admit direct access by root identifier.
All tests use a shared failure flag with common writes of one.
Only an arc identifier is arbitrated at a selection cell, and
its endpoints and witness are read from that record.
The certificate test uses $O(M)$ processors, constant private memory,
and constant time. It performs no counting or compaction. A failed test
changes only temporary arrays. After acceptance, the root links join
the existing parent trees by stars. Those trees may still have depth
greater than one; terminal pointer shortcuts flatten them before
recovery starts. Each flatness test uses a cleared shared flag and
common writes of one, so it takes constant time with $O(q)$ processors.

\begin{lemma}[Termination and terminal flattening]
\label{lem:star-close}
An accepted certificate selects one witnessed star per current root
component. On $\mathcal E$, the certificate is accepted after
$R=O(\log d+\log\log n)$ forward rounds.
Its terminal flattening takes $O(\log(R+1))$ additional time and witness
layers, with $O(M)$ processors. Expansion and terminal flattening
together take $O(R)$ time.
\end{lemma}
\begin{proof}
In \algstep{alg:expand}{step:close-proposals}, each noncenter root
chooses an incident arc to a center; every center labels itself.
The comparisons in \algstep{alg:expand}{step:close-check} make the
labels constant along every root edge. Hence each root component has
one center and every other root chooses an edge to it. A second center
would retain a different label and fail the test. Since all current
nonloop edges have root endpoints, joining these stars connects exactly
the parent trees in each component of $G_0$ by
Lemma~\ref{lem:forward-invariants}. No flatness assumption is needed.

On a clique, only its smallest root identifier survives
\algstep{alg:expand}{step:close-centers}; every proposal exists and all
consistency checks pass. Failed tests leave the forward state unchanged.
Lemma~\ref{lem:hash-schedule} supplies such a simultaneous clique
boundary within the asserted number of rounds, proving acceptance.

Before the terminal root links, every retained vertex has level between
$1$ and $R+1$, and levels strictly increase along parent arcs.
Thus each parent tree has height at most $R$. The accepted star links
add at most one edge to any root path, giving height at most $R+1$.
After $a$ synchronous shortcuts, that height is at most
$\lceil(R+1)/2^a\rceil$. Consequently
\algstep{alg:expand}{step:close-flatten} finishes in
$O(\log(R+1))$ iterations. Each iteration uses the same $O(q)$ vertex
processors and adds one binary witness layer. This occurs only once,
after expansion has stopped. Since $\log(R+1)=O(R)$, the total time
and number of witness layers remain $O(R)$.
No further expansion is executed after acceptance. The terminal root
links need not preserve the forward levels: subsequent shortcuts use
only the parent-forest structure, and inverse recovery uses its separate
rank invariant.
\end{proof}

\subsection{Virtual edges and immutable witnesses}
\label{sec:sf-witnesses}
All witnesses represent walks in the base graph \(G_0\) left by
workspace preparation. Each current directed arc stores the
identifier of an immutable witness record.
Write \(W(C)\) for the walk represented by a record \(C\).
A base record \(\mathsf{BASE}(u,v,e)\) stores its current endpoints
\(u,v\) in \(G_0\) and its original input-arc identifier \(e\).
A compound record references two earlier witnesses with matching
endpoints, where \(\mathsf{tail}\) and \(\mathsf{head}\) denote their
start and end vertices:
\[
 \mathsf{CONCAT}(u,v;A,B),
 \quad
 \mathsf{head}(A)=\mathsf{tail}(B).
\]
An identity record \(\mathsf{LOOP}(u)\) represents a zero-length walk.
Each reference is one word encoding a record identifier and a direction
bit. Reversing a reference flips that bit; it does not copy a walk.
A compound record uses $O(1)$ shared words: two endpoints, two oriented
child references, and its creation-layer number. Arbitrary writes
publish the encoded reference as one word, so a winner's endpoints and
children are always read from the same immutable record. It represents the walk
\[
 W(\mathsf{CONCAT}(u,v;A,B))=W(A)\circ W(B).
\]
These walks are in \(G_0\). Original endpoints are read from the
immutable input array only when their original endpoints are recovered.

Each forward operation has a short witness in the preceding state,
as listed in Table~\ref{tab:forward-witnesses}.
\begin{table}[htbp]
\centering
\caption{Witness walks for the forward operations.}
\label{tab:forward-witnesses}
\smallskip
\renewcommand{\arraystretch}{1.25}
\setlength{\tabcolsep}{10pt}
\begin{tabular}{|l|l|}
\hline
\textbf{New arc} & \textbf{Witness walk}\\
\hline
Expansion \(v\to u\) & \(v\to w\to u\)\\
\hline
\textsc{Maxlink}: \(v\to p[w]\) & \(v\to w\to p[w]\)\\
\hline
\textsc{Shortcut}: \(v\to p[p[v]]\) & \(v\to p[v]\to p[p[v]]\)\\
\hline
\textsc{Alter}: \((p[v],p[w])\) & \(p[v]\to v\to w\to p[w]\)\\
\hline
\end{tabular}
\end{table}
Each displayed arc may itself have a compound witness.
Its full walk is represented by the referenced record and is never
copied. The first and last arcs in the \textsc{Alter} witness use
the endpoint-parent records, with the first read in reverse.
We encode this three-arc walk by two binary concatenations.
Thus every primitive operation contributes only \(O(1)\) witness
layers.

Number binary creation layers consecutively, with base arcs and
identity records at layer zero. A record created at layer $j$ may
reference only records created before $j$. In particular, the two
\textsc{Maxlink} iterations occupy separate layers, and the two
concatenations implementing \textsc{Alter} occupy consecutive layers.
Neighbor-table reads finish before a new expansion layer is published.
Every terminal shortcut also receives its own layer. Let $S$ be the
last layer number. Each round has only a constant number of layers;
terminal flattening contributes $O(\log(R+1))$ further layers.

\paragraph{Reading one inverse layer.}
Write $\overline C$ for a reference with its direction reversed and
$\tau(C)$ for the stored creation layer. To split a reference $C$ at
layer $j$, use its stored children when $\tau(C)=j$. If its forward
children are $(A,B)$, the children of $\overline C$ are
$(\overline B,\overline A)$, in that order. If $\tau(C)<j$, instead
use $(C,\mathsf{LOOP}(\mathsf{head}(C)))$. This padding pair has the
same endpoints and uses only records from layers at most $j-1$.
It is interpreted locally and requires no newly allocated record.
Thus a split reads a constant number of words, even if the represented
walk is long or its creation layer is much earlier. A reference with
$\tau(C)>j$ is forbidden by the recovery invariant below.

For analysis, let $\mathcal W_j$ be the undirected graph on $V(G_0)$
containing the endpoint pair of every record created by layer $j$,
including all base arcs. This graph is not constructed or scanned.
Each record represents a walk in $G_0$, so $\mathcal W_j$ has exactly
the components of $G_0$ for every $j$.

Each retained vertex \(v\) has a shared pointer
\(\mathsf{curWit}[v]\) to the witness of its current parent arc
\(v\to p[v]\).
Let \(F_S\) denote the final parent forest.
The shared nonnegative integer rank \(r[v]\) is used only during
inverse recovery.
After inverse recovery, the attachment vertex of each preparation
component is the original tail of its outgoing selected arc, or its
retained representative if it has no outgoing arc.

\noindent\begin{minipage}{\linewidth}
\begin{framed}
\small
\setlength{\parindent}{0pt}
\algcaption{Witnessed Spanning Forest}
\label{alg:additive-sf}\par
\smallskip
\textbf{Input:}
An undirected graph \(G\) with immutable original edge identifiers.
\par
\textbf{Output:}
An original-edge spanning forest and its parent-edge array, or
\textsc{Failure} if preparation or a forward resource check fails.
\par
\smallskip
\begin{enumerate}[label=\arabic*.,ref=\arabic*,leftmargin=1.5em,topsep=3pt,itemsep=3pt,parsep=0pt]
\item \label{step:sf-prepare}
Prepare \(G_0\) and the records of \(F_0\) as in
\S\ref{sec:hash-schedule}.
Return \textsc{Failure} if preparation fails.
Initialize base witnesses for retained arcs and
\(\mathsf{curWit}[v]\gets\mathsf{LOOP}(v)\).
If no representative remains active, orient \(F_0\) with
Algorithm~\ref{alg:restore} and return.

\item \label{step:sf-middle}
Repeat Algorithm~\ref{alg:expand} until \textsc{Finished},
recording witnesses for every arc and parent update.
If it reports \textsc{Failure}, return \textsc{Failure} immediately.

\item \label{step:sf-finish}
Retain the final parents and witnesses as the virtual star forest \(F_S\).

\item \label{step:sf-initialize-rank}
Set \(r[v]\gets[p[v]=v]\) for each retained vertex.
Prepare the grouped-maximum workspace and bit-power table once.

\item \label{step:sf-reverse}
For \(j=S,S-1,\ldots,1\), call
Algorithm~\ref{alg:sf-inverse} on the current parent witnesses and
ranks.

\item \label{step:sf-original-parents}
Recover the original endpoints of selected base arcs.
Run Algorithm~\ref{alg:restore} with the attachment vertices as roots.
Combine internal and outgoing parents; return the parent and
parent-edge arrays.
\end{enumerate}
\end{framed}
\end{minipage}\par\smallskip

To recover the original endpoints of a selected base arc \((u,v)\),
read its immutable original
endpoints \((a,b)\). Exchange them if the last preparation label of
\(a\) is not \(u\). Then \(a\) is the attachment vertex in the
component of \(u\), and its outgoing parent is \(b\).
Each recovered base vertex has at most one outgoing parent arc, so
these attachment choices are unique. The preparation recovery uses
separate temporary parent arrays before the outgoing arcs are added.
For every component already completed during preparation, keep its
final representative as its designated root. These components are
recovered in the same parallel call; they require no expansion or
inverse witness layers.

Each witness layer reserves a constant number of shared record slots
per processor. Records remain accessible through their identifiers
after the layer finishes.
For expansion, assign one processor to each pair of neighbor-table
slots within a root's allocated block. For slots \(a,b\) at root \(u\),
it reads \((w,A)\) from \(H(u)[a]\) and \((v,B)\) from \(H(w)[b]\).
It writes \(\mathsf{CONCAT}(u,v;A,B)\) to its reserved record slot.
Only the record identifier is written to the hash cell.
The hash procedure either rejects a candidate or selects one identifier.
All fields are then read from that record.
On \(\mathcal E\), the allocated root blocks use \(O(M)\) processors
in total. Each layer therefore has \(O(M)\) record slots.
Each primitive allocates records only for the constant number of
candidates generated by one of its assigned processors. This includes
parent updates and both binary stages of \textsc{Alter}, as well as
two-hop expansion. A retained current arc stores a reference to a
winning record; losing candidates need no current arc slot.
Record identifiers address reserved shared slots directly, and each
record stores its creation layer. Immutable storage is never reclaimed
while a reference may still reach it. Reading a reference therefore
takes constant time regardless of its age or how many other records
share its children.

The workspace bound \(q\log^2 n=O(M)\) gives a processor
surplus to assign one processor per candidate, key bit, and hash
replica in the grouped-maximum procedure.

Neighbor expansion leaves \(\mathsf{curWit}\) unchanged.
When \textsc{Maxlink} chooses \(v\to p[w]\), it
concatenates the witness for \(v\to w\) with the current parent
witness of \(w\). The new identifier is stored in
\(\mathsf{curWit}[v]\).
\textsc{Shortcut} concatenates the current witnesses of
\(v\to p[v]\) and \(p[v]\to p[p[v]]\).
Every witness is assembled from the old shared arrays before those
arrays are overwritten. The parent and its new witness reference
are then published together. At termination, the current references
of the final star forest are the starting records for recovery.
Earlier records remain immutable and are reached through their child
references.

\begin{lemma}[Witness soundness]
\label{lem:sf-witness}
Each witness \(C\) represents a valid walk \(W(C)\) between its stored
endpoints. In particular, every parent witness saved by
Algorithm~\ref{alg:additive-sf} connects the two corresponding
components.
\end{lemma}
\begin{proof}
We induct on the creation order of the records.
The base records initialized in \algstep{alg:additive-sf}{step:sf-prepare}
represent arcs of \(G_0\). For \(C=(u,v;A,B)\), the construction checks that
the head of \(A\) is the tail of \(B\). By induction, \(W(A)\)
and \(W(B)\) are valid walks with these endpoints. Their concatenation
is a walk from \(u\) to \(v\). A parent update stores either an
existing witness or a concatenation of valid witnesses.
The updates recorded in \algstep{alg:additive-sf}{step:sf-middle}
use exactly the walks listed above.
The terminal root links in
\algstep{alg:expand}{step:close-finish} copy existing current-arc
witnesses. Each terminal shortcut concatenates two valid parent
witnesses from its input snapshot. Thus every parent of the final
star forest has a valid witness.
\end{proof}

\begin{lemma}[The final virtual star forest]
\label{lem:sf-virtual-height}
The final parent array \(F_S\) is a forest on \(V(G_0)\) with the
same components as \(G_0\), one center per component, and rooted
depth at most one. Every parent arc has a witness in \(G_0\).
\end{lemma}
\begin{proof}
Let \(p_{\rm old}\) denote the parent array when the termination
certificate passes. All current nonloop arcs have root endpoints
by \algstep{alg:expand}{step:close-flat}; the parent trees need not
yet be flat.
Lemma~\ref{lem:forward-invariants} implies that the current root graph
connects exactly the parent trees belonging to each component of \(G_0\).
Lemma~\ref{lem:star-close} assigns one center \(c\) to its roots.
Installing the selected root arcs joins these disjoint parent trees
by a star, so it introduces no cycle and gives one tree per component.
Every retained vertex has its old parent path to a root, followed
by at most one selected root edge to $c$. Repeated shortcuts in
\algstep{alg:expand}{step:close-flatten} preserve this root and
eventually make every vertex point directly to it. Each shortcut
concatenates two witnesses from its own unchanged input snapshot;
it does not copy or enumerate their underlying walks.
Lemma~\ref{lem:sf-witness} makes these witnesses valid.
Thus \(F_S\) spans the same components and has depth at most one.
\end{proof}

\subsection{Rank-preserving inverse expansion}
\label{sec:sf-inverse}
In this subsection, \(p\) denotes the parent map of the forest being
recovered, rather than the forward connectivity state.
For witness layer \(j\), let \(F_j\) be the current recovery forest.
Suppose a current virtual parent arc \(v\to p(v)\) has witness
\[
 v\longrightarrow w_v\longrightarrow p(v).
\]
For every nonroot \(v\), send the candidate
\[
 (w_v,\,2r(p(v))-1,\,v)
\]
to \(w_v\). At \(x\), select the maximum candidate key together with
the default key \(2r(x)\). If a candidate wins, choose its second
arc as the new parent arc of \(x\). If the default wins and \(x\)
is a nonroot, choose its first arc \(x\to w_x\). Roots keep their
self-pointers. Resolve ties arbitrarily.

A candidate tuple lists its \(\mathsf{group}\), \(\mathsf{key}\), and
\(\mathsf{payload}\): the target slot, proposed rank, and originating
vertex. A default candidate has payload \(\nil\).
The shared cell \(\mathsf{win}[g]\) stores the winning record identifier
for group \(g\).
Use \(2q\) candidate slots and key width \(w=S+2\);
the rank bound below shows that this width suffices for all layers.

\noindent\begin{minipage}{\linewidth}
\begin{framed}
\small
\setlength{\parindent}{0pt}
\algcaption{Inverse Witness Expansion}
\label{alg:sf-inverse}\par
\smallskip
\textbf{Input:} Layer \(j\), recovery forest \(p\), parent witnesses,
and ranks \(r\).\par
\textbf{Output:} A parent forest using witnesses from earlier layers.\par
\smallskip
\begin{enumerate}[label=\arabic*.,ref=\arabic*,leftmargin=1.5em,topsep=3pt,itemsep=3pt,parsep=0pt]
\item \label{step:inverse-split}
For each nonroot $v$, split its current reference at layer $j$,
respecting its direction and creation layer, into
$A_v:v\to w_v$ and $B_v:w_v\to p[v]$.
Store these references and $w_v$ in shared vertex-indexed arrays.

\item \label{step:inverse-candidates}
Clear the \(2q\) candidate slots.
Each vertex \(x\) creates a default and each nonroot \(v\) an
additional candidate:
\[
 (g(x),2r[x],\nil),\qquad
 (g(w_v),2r[p[v]]-1,v).
\]
Mark created records valid and call Algorithm~\ref{alg:group-max}.

\item \label{step:inverse-select}
At each \(x\), read \(s\gets\mathsf{win}[g(x)]\) and set
\(r_{\rm new}[x]\gets\mathsf{key}[s]\).
If its payload is \(v\ne\nil\), set \(p_{\rm new}[x]\gets p[v]\)
and select \(B_v\).
Otherwise set \(p_{\rm new}[x]\gets w_x\) and select \(A_x\) for
nonroots; keep root self-pointers and identity witnesses.

\item \label{step:inverse-publish}
Replace the parent, witness, and rank arrays by the new arrays.
\end{enumerate}
\end{framed}
\end{minipage}
\par\smallskip

\paragraph{Computing the new ranks in constant time.}
Let \(g(x)\in\{0,\ldots,q-1\}\) be the compacted slot stored for
representative \(x\). Candidate records occupy two fixed shared
slots per representative: the default for \(x\) uses slot \(2g(x)\),
and the additional candidate from a nonroot \(v\) uses slot
\(2g(v)+1\). A root's additional slot is inactive.
The fields \(\mathsf{group}\), \(\mathsf{key}\), and
\(\mathsf{payload}\) are separate shared arrays indexed by record
slot, each with address equal to its base plus the slot number.
The stored map \(g\) and all winner cells also admit direct access.

The records in group \(g(x)\) have keys exactly
\[
 \{2r(x)\}\ \cup\
 \{\,2r(p(v))-1:\ v\text{ is a nonroot and }w_v=x\,\}.
\]
Thus the value \(r'(x)\) is the key of the record returned for this
group, read in \algstep{alg:sf-inverse}{step:inverse-select}.
There is one grouped-maximum call for all vertices, with at most
\(2q\) records in total.
Algorithm~\ref{alg:group-max} assigns one processor per record,
key bit, and hash replica. With \(w=O(\log n)\) key bits and
\(L=O(\log n)\) replicas, both its processor count and its shared
workspace are \(O(qwL+q)=O(q\log^2 n)=O(M)\).
The routine verifies its provisional winners and retries with fresh
hashes if needed. Its accepted result is exact for every write
resolution, and a call takes \(O(1)\) time with high probability,
even if one group contains \(\Theta(q)\) records.
Only after the call returns does each vertex read its winning key
and publish its new rank and parent. Failed attempts leave all input
ranks, witnesses, and candidates unchanged.

\begin{lemma}[Rank-preserving inverse expansion]
\label{lem:sf-inverse-rank}
Let $F_j$ span $\mathcal W_j$, with each parent reference created
at a layer at most $j$. If its nonnegative integer ranks strictly
increase along nonroot parent arcs, Algorithm~\ref{alg:sf-inverse}
produces a spanning forest $F_{j-1}$ of $\mathcal W_{j-1}$ with the same
roots. Every new parent reference was created at a layer at most $j-1$. The new rank
\[
 r'(x)=\max\bigl(2r(x),
        \max_{v:w_v=x}(2r(p(v))-1)\bigr)
\]
strictly increases along every nonroot parent arc of \(F_{j-1}\), and
\(\max_x r'(x)\le2\max_x r(x)\).
\end{lemma}
\begin{proof}
The split in \algstep{alg:sf-inverse}{step:inverse-split} gives
two correctly oriented child references from layers at most $j-1$.
This holds for a record born at $j$ by the strict creation order, and
for an older record by the identity-padding rule. Hence every selected
child is an edge of $\mathcal W_{j-1}$ with the asserted endpoints.
The candidates in \algstep{alg:sf-inverse}{step:inverse-candidates}
give exactly the displayed formula for $r'$. The grouped-maximum call verifies
its winners before returning, so the formula is exact on every
completed layer, independently of the hash values.
Consider the parent chosen in
\algstep{alg:sf-inverse}{step:inverse-select}.
If \(x\) receives a winning second child arc from \(v\), then
\[
 r'(x)=2r(p(v))-1<2r(p(v))\le r'(p(v)).
\]
If \(x\) keeps its first child arc to \(w_x\), then the old rank
inequality \(r(p(x))\ge r(x)+1\) gives
\[
 r'(x)=2r(x)<2r(p(x))-1\le r'(w_x).
\]
Every selected parent arc therefore increases \(r'\), so the new
parent graph is acyclic. An intermediate witness vertex lies in the
same component as the endpoints. Since \(F_j\) spans that component,
all three vertices lie in one tree of \(F_j\).
An old root \(z\) has maximum rank in its tree. Any candidate sent to
\(z\) has key at most \(2r(z)-1\), so its default key wins.
Every old nonroot receives a parent of larger rank. Hence the roots
are unchanged. No selected arc leaves an old component. Following new
parents must end at a root, and each old component has exactly one such
root. Thus the new forest spans exactly the same components. Since
$\mathcal W_j$ and $\mathcal W_{j-1}$ have the components of $G_0$,
this also establishes the spanning invariant for the next layer.
A self-loop child cannot be selected for a nonroot: either of the
strict rank inequalities would then contradict itself.
Every default key is at most \(2\max_x r(x)\), and every other
key is strictly smaller than this bound. Thus the maximum rank
at most doubles.
\end{proof}

\paragraph{Completion and processor reuse.}
The final star forest spans $\mathcal W_S$, and its ranks zero and one
satisfy the inverse invariant. Applying
Lemma~\ref{lem:sf-inverse-rank} for $j=S,\ldots,1$ therefore leaves
only layer-zero references. A nonroot reference is then an original
base arc, rather than an identity loop. At each layer only the one
current parent reference of each retained vertex is inspected. Shared
subrecords are read concurrently, never expanded into separate lists.
There are at most $q$ defaults and $q$ nonroot candidates; the same
candidate slots and maximum workspace are reused at the next layer.
The algorithm never scans all $O(MS)$ historical records in a layer
and never assigns a processor to each edge in a represented walk.

\paragraph{Hashing the prefix keys.}
The input has \(K\) fixed record slots, indexed by \(s=0,\ldots,K-1\),
and at most \(q\) groups.
Slot \(s\) stores a validity bit, a group identifier
\(g_s=\mathsf{group}[s]\in\{0,\ldots,q-1\}\), a \(w\)-bit key
\(a_s=\mathsf{key}[s]\), and its payload.
Group \(g(x)\) contains the candidates whose target is \(x\).
If \(K=0\), return empty group winners directly.
Otherwise assume \(w\ge1\).
Use \(L=\Theta(\log n)\) independent hash tables, each of capacity
\(B\), where \(B\) is the smallest power of two at least \(4Kw\).
For a desired failure exponent \(\eta>0\), choose the constant in
\(L\) sufficiently large for the bound in Lemma~\ref{lem:max}.
For a record with target group \(g\) and key \(a\), let
\(j\in\{1,\ldots,w\}\) be a prefix length and encode its prefix by
\[
 \kappa(g,j,b)=g\,2^{w+1}+2^j+b,
 \qquad b=\lfloor a/2^{w-j}\rfloor .
\]
The intervals for different \(j\) are disjoint, so this encoding is
injective. These integers are keys, not shared-memory addresses.
When \(b\) is even, \(\kappa(g,j,b)+1\) encodes the larger sibling
prefix obtained by changing the last bit from zero to one.

Let \(\ell=O(\log n)\) be the word width, large enough for the
encoded keys, \(B\), and all workspace addresses.
For each table index \(t\in\{0,\ldots,L-1\}\), independently sample
an odd \(\ell\)-bit multiplier \(\alpha_t\).
For an encoded key \(k\), use the multiply-shift hash
\[
 h_t(k)=\left\lfloor
       \frac{\alpha_t k\bmod 2^\ell}{2^\ell/B}
       \right\rfloor,
 \qquad
 \Pr[h_t(k)=h_t(k')]\le\frac{2}{B}\quad(k\ne k').
\]
The hash seed uses a constant number of words. Multiplication and
shifts evaluate the hash in constant time~\cite{DHKP97}.
Only the multipliers for different replicas are independent;
the collision bound above is all that is needed within one replica.
The shared flag \(\mathsf{reject}[s]\) marks an eliminated candidate,
and \(\mathsf{win}[g]\) stores the selected record identifier for
group \(g\). The shared flag \(\mathsf{bad}\) records a failed
verification.

\noindent\begin{minipage}{\linewidth}
\begin{framed}
\small
\setlength{\parindent}{0pt}
\algcaption{Verified Grouped Maximum}
\label{alg:group-max}\par
\smallskip
\textbf{Input:} \(K\) fixed shared candidate slots, each storing
a validity bit, a target group, a \(w\)-bit key, and a payload.\par
\textbf{Output:} One exact maximizing record identifier for every
nonempty group.\par
\smallskip
\textbf{Attempt:} Repeat until verification succeeds.\par
\begin{enumerate}[label=\arabic*.,ref=\arabic*,leftmargin=1.5em,topsep=3pt,itemsep=3pt,parsep=0pt]
\item \label{step:group-clear}
Clear the tables and flags; set group winners to \(\nil\).
Sample fresh independent multipliers \(\alpha_t\).

\item \label{step:group-prefix}
In parallel for each valid slot \(s\), prefix length \(j\), and
replica \(t\), compute
\(b\gets\lfloor a_s/2^{w-j}\rfloor\),
\(k\gets\kappa(g_s,j,b)\).

\item \label{step:group-occupy}
Each active processor writes \(k\) to \(\mathsf{Tab}[t,h_t(k)]\).

\item \label{step:group-reject}
Each active processor with even \(b\) sets \(\mathsf{reject}[s]\gets1\)
if \(\mathsf{Tab}[t,h_t(k+1)]=k+1\).

\item \label{step:group-publish}
Each valid unrejected record \(s\) writes \(s\) to
\(\mathsf{win}[g_s]\).

\item \label{step:group-verify}
Each valid record \(s\) reads \(u\gets\mathsf{win}[g_s]\).
If \(u=\nil\), write one to \(\mathsf{bad}\).
Otherwise, if \(a_s>\mathsf{key}[u]\), write one to \(\mathsf{bad}\).

\item \label{step:group-return}
Return \(\mathsf{win}\) if \(\mathsf{bad}=0\); otherwise retry.
\end{enumerate}
\end{framed}
\end{minipage}\par\smallskip

\paragraph{Shared storage and preparation.}
The table is a shared \(L\)-by-\(B\) array:
\[
 \mathsf{addr}_{\mathsf{Tab}}(t,z)
   =\mathsf{base}_{\mathsf{Tab}}+tB+z.
\]
It stores one complete encoded key per cell; zero denotes an empty
cell since every encoded key is positive.
Assign one processor to each record slot, prefix length, and table
replica. These fixed index ranges determine the assignment directly.
Validity checks skip empty input slots. One processor per record
and per group handles publishing and verification.
Table initialization uses one processor per cell; this is only
a constant-factor larger pool since \(B=O(Kw)\).
The input snapshot is fixed before sampling the multipliers.
All candidate fields remain read-only during attempts.

In inverse recovery set \(K=2q\) and \(w=S+2\), where \(S\)
is the number of witness layers.
The final star forest starts with ranks zero and one in
\algstep{alg:additive-sf}{step:sf-initialize-rank}.
Each accepted inverse layer at most doubles their maximum,
so the largest key in all \(S\) layers is at most \(2^S\).
Thus every candidate fits in \(w\) bits, and \(w=O(\log n)\)
on the forward event.
Prepare the shared powers \(2^j\), \(0\le j<\ell\), once by word
shifts, using one processor per exponent.
To choose \(B\), one processor per positive exponent \(j<\ell\)
tests \(2^{j-1}<4Kw\le2^j\); the unique successful processor
writes \(B=2^j\). This setup takes \(O(1)\) time with \(O(\log n)\)
processors. Every prefix is then computed by a single word shift.

The table, rejection flags, winner cells, seeds, and bit powers
occupy \(O(KwL+q)=O(q\log^2 n)=O(M)\) shared words.
All attempts and inverse layers reuse this workspace.
No processor count depends on a group's number of incoming candidates,
and no list of prefix keys is compacted.

\begin{lemma}[Verified grouped maximum]
\label{lem:max}
For \(K\) fixed record slots with \(w\)-bit keys and \(q\) groups,
where \(w=O(\log n)\) and \(K,q\le\poly(n)\),
Algorithm~\ref{alg:group-max} returns exact group maxima whenever
it returns, for every permitted concurrent-write resolution.
An attempt takes \(O(1)\) time with \(O(KwL+q)\) processors and
shared words. Its probability of rejection is at most \(Kw\,2^{-L}\).
For polynomially bounded \(K,q\), \(w=O(\log n)\), and any fixed
\(\eta>0\), a sufficiently large constant in
\(L=\Theta(\log n)\) makes the first attempt succeed with
probability at least \(1-n^{-\eta-2}\).
The algorithm terminates almost surely.
\end{lemma}
\begin{proof}
Let \(D\) be the set of distinct encoded keys inserted in
\algstep{alg:group-max}{step:group-occupy}; \(|D|\le Kw\).
Every occupied cell contains a key in \(D\), regardless of its
winning writer. Consequently the exact-key test in
\algstep{alg:group-max}{step:group-reject} cannot report an absent
prefix. A maximum candidate has no larger sibling prefix in its
group, so it is never rejected. Every nonempty group therefore has
a provisional winner. Verification in
\algstep{alg:group-max}{step:group-verify} compares that winner with
every input key in the group. A return in
\algstep{alg:group-max}{step:group-return} is possible exactly when
all provisional winners are maxima. This proves unconditional
correctness of every accepted result, including ties and duplicates.

Fix an inserted key \(k\in D\).
In one replica the probability that a different key shares its bucket
is at most
\[
 \sum_{k'\in D\setminus\{k\}}\Pr[h_t(k')=h_t(k)]
 \le \frac{2(|D|-1)}{B}<\frac12.
\]
Independent multipliers imply that \(k\) has no collision-free
replica with probability at most \(2^{-L}\).
If every key has such a replica, all membership queries for keys
in \(D\) succeed. For any nonmaximum candidate, comparison with a
larger key at the first differing bit gives an inserted larger
sibling prefix. That candidate is then rejected, so verification
passes for every possible write resolution.
A union bound over \(D\) bounds rejection by \(Kw\,2^{-L}\).
Choosing the constant in \(L\) sufficiently large gives the stated
probability.
This proof uses no independence between keys within a replica.

Each stage has constant depth. There are \(KwL\) prefix processors,
\(BL=O(KwL)\) table cells to initialize, and \(O(K+q)\) other
cells and processors. Every processor stores only \(O(1)\) words.
Fresh multipliers are independent of the fixed input and all previous
attempts. The conditional rejection probability of each attempt is
at most \(Kw\,2^{-L}\le1/2\), so the probability of more than \(a\)
attempts is at most \(2^{-a}\).
This also proves almost-sure termination.
\end{proof}

\begin{corollary}[All adaptive inverse layers]
\label{cor:inverse-hashing}
Fix a forward trace satisfying \(\mathcal E\).
For every fixed \(\eta>0\), with conditional probability at least
\(1-n^{-\eta}\), all \(S=O(\log n)\) inverse layers finish their
grouped-maximum calls on the first attempt.
Denote this event by \(\mathcal H\).
Their total time is \(O(S)\), with \(O(M)\) processors and
\(O(M)\) additional shared workspace.
Every accepted layer is correct even outside \(\mathcal H\).
\end{corollary}
\begin{proof}
Condition on the entire execution before any one call, including the
earlier inverse winners and all previous random bits.
Its candidates are now fixed; its fresh multipliers remain independent.
Lemma~\ref{lem:max} therefore bounds its conditional rejection
probability by \(n^{-\eta-2}\), even though later candidates are
chosen adaptively. A union bound over \(S=O(\log n)\) calls gives
the stated bound. In this application \(K=2q\), \(w,L=O(\log n)\),
and preparation guarantees \(q\log^2 n=O(M)\).
The resource bounds follow, with one shared workspace reused by all
layers. Algorithm~\ref{alg:sf-inverse} updates parents and ranks
only after verification has returned exact maxima.
\end{proof}

\subsection{Running time and output depth}
\label{sec:sf-depth}
The inverse construction controls depth through ranks.
Recovery starts from the final star forest, whose ranks are zero
and one. Each inverse layer at most doubles the maximum rank,
so reversing all \(S\) layers gives depth at most \(2^S\).
The preparation forest needs a separate diameter bound, since flat
temporary labels do not imply shallow original-edge trees.
The following elementary recurrence handles that distinction.
Here \(\Delta_i\) is the maximum diameter before a joining phase,
and \(h_i\) is the height of its connection forest.

\begin{lemma}[Diameter after joining tree components]
\label{lem:diamrec}
If tree components of diameter at most \(\Delta_i\) are joined along a
connection tree of height at most \(h_i\), then
\[
 \Delta_{i+1}+1\le(2h_i+1)(\Delta_i+1).
\]
\end{lemma}
\begin{proof}
Between any two contracted vertices there are at most \(2h_i\)
connection edges. The corresponding original-tree path uses at most
\(2h_i\) connecting edges and at most \(2h_i+1\) old components.
Each segment inside an old component has length at most \(\Delta_i\).
This includes the first and last components on the path. The total
length is therefore at most \((2h_i+1)\Delta_i+2h_i\).
Rearranging gives the claim. The connection structure contains exactly
the selected forest edges, so parallel input edges introduce no cycle.
\end{proof}

\begin{theorem}[Additive-time spanning forest]
\label{thm:sf}
On \(\mathcal E\cap\mathcal H\), Algorithm~\ref{alg:additive-sf}
computes a spanning forest of original edges in
\[
 O(\log d+\log\log n)
\]
time using \(O(m+n)\) ARBITRARY CRCW processors and \(O(1)\) private
words per processor. Its output forest has maximum rooted depth
\[
 h_{\rm SF}\le 2^{O(\log d+\log\log n)}.
\]
The grouped-maximum workspace is \(O(m+n)\) shared words.
The combined event holds with good probability; conditional on
every forward trace in \(\mathcal E\), the event \(\mathcal H\)
has probability at least \(1-n^{-\eta}\).
Every completed recovery from a trace in \(\mathcal E\) is correct
and has the stated depth bound, including executions outside
\(\mathcal H\).
\end{theorem}
\begin{proof}
If preparation leaves no active representative, \(F_0\) already
spans the input. Lemma~\ref{lem:diamrec} bounds its diameter by
\(3^{R_0}-1\), and its star-contraction history is recovered in
\(O(R_0)\) time by Algorithm~\ref{alg:restore}. The claims follow.
Assume henceforth that at least one representative remains active.
Workspace preparation, including the shared budget tables, takes
$O(\log\log n)$ time once, by Lemma~\ref{lem:hash-schedule}.

By Lemma~\ref{lem:star-close}, the number of expansion rounds is
$R=O(\log d+\log\log n)$. Each forward round has a constant
number of binary witness layers. Terminal flattening adds
$O(\log(R+1))$ layers, so
\[
 S=O(R+\log(R+1))=O(\log d+\log\log n).
\]
In particular, flattening contributes an additive cost, not a cost
paid at every expansion round. Each forward operation creates \(O(1)\) witness
records per active processor. Recording uses the same \(O(M)\)
processor pool. Lemma~\ref{lem:sf-virtual-height} gives the final star
forest \(F_S\) of depth at most one.
Its ranks are initialized by one parallel root test.
The shared bit powers and hashed-maximum workspace are configured
once in \(O(1)\) time.

Each inverse layer creates one candidate per parent arc, besides
the default candidates, so there are at most \(2q\) record slots.
Corollary~\ref{cor:inverse-hashing} gives \(O(1)\) time per layer
on \(\mathcal H\), with \(O(M)\) processors and shared workspace.
Verification makes the returned group maxima exact.
Lemma~\ref{lem:sf-inverse-rank} therefore preserves the roots and connected
components on every completed layer.
All \(S\) inverse layers take \(O(S)\) time on \(\mathcal H\)
and recover the base-edge forest by the creation-layer invariant.

Algorithm~\ref{alg:restore} roots each preparation tree at its
designated attachment vertex. Replacing every base vertex by this
tree and retaining the recovered base edges preserves acyclicity
and connects exactly the original input components.
We next bound the depth of this original-edge forest.
The workspace preparation in
\algstep{alg:additive-sf}{step:sf-prepare} has
\(R_0=O(\log\log n)\) phases, each with star-shaped connection trees.
Lemma~\ref{lem:diamrec} bounds the diameter of each preparation
tree by \(3^{R_0}-1\).
Let \(h_V\) be the depth of the recovered forest on \(G_0\).
The initial ranks in
\algstep{alg:additive-sf}{step:sf-initialize-rank} are at most
\(1\). Algorithm~\ref{alg:sf-inverse} at most doubles the
largest rank, and every parent arc increases rank by at least one.
Consequently \(h_V\le2^S\).
The preparation recovery directs each internal tree toward its outgoing
base-edge attachment, or toward its designated root if there is no
outgoing edge. Hence a root path in the output forest follows a
parent path of the recovered base forest. Once it leaves a preparation
tree along that tree's unique outgoing base edge, it cannot revisit it,
since the base parent forest is acyclic. It uses at most \(h_V\)
base edges and at most \(h_V+1\) preparation trees.
Each internal segment has length at most \(3^{R_0}-1\), so
\[
 h_{\rm SF}+1
 \le 3^{R_0}(h_V+1)
 \le 3^{R_0}(2^S+1).
\]
Since \(R_0=O(\log\log n)\) and \(S=O(\log d+\log\log n)\), this gives
\(h_{\rm SF}\le2^{O(\log d+\log\log n)}\).
The call to Algorithm~\ref{alg:restore} in
\algstep{alg:additive-sf}{step:sf-original-parents} reverses only
the preparation phases. Their connection trees have height one,
so this call takes \(O(R_0)=O(\log\log n)\) additional time.
The total time is therefore $O(\log\log n+R+S+R_0)$, which is
$O(\log d+\log\log n)$ on $\mathcal E\cap\mathcal H$.
We reuse the processor pool between layers. There are $O(M)$ current
arc slots by Lemma~\ref{lem:hash-schedule}, $O(MS)$ immutable witness
words, and $O(nR_0)$ preparation-history words. These are shared arrays,
not private memory or simultaneously active processor requirements.
The grouped-maximum workspace is reused at each inverse layer.
The algorithm therefore uses $O(M)$ processors throughout and
$O(M(S+R_0))$ shared words, in addition to its input and output.
The event $\mathcal E$ has the stated good probability. Conditional on
each such forward trace, Corollary~\ref{cor:inverse-hashing} gives
$\Pr[\mathcal H\mid\text{trace}]\ge1-n^{-\eta}$; averaging and a
union bound preserve good probability for their intersection.
Choose $\eta$ large enough for the polynomial input-size convention.
Together with the depth calculation, this proves
Theorem~\ref{thm:main-sf}.
\end{proof}

\section{2-Edge Connectivity without DFS Sequence}
\label{sec:bridge}
In this section, we introduce an algorithm to find bridges without constructing a DFS sequence. We first apply Algorithm \ref{alg:additive-sf} to compute a rooted spanning forest and its parent array. Only a tree edge can be a bridge, and a tree edge is a bridge if and only if it lies on no fundamental cycle formed by a non-tree edge. Therefore we find bridges by marking the tree edges covered by fundamental cycles and the remaining unmarked tree edges are bridges.

\begin{definition}[Bridge]
\label{def:bridge}
An edge $e$ of an undirected graph $G$ is a bridge if deleting $e$
increases the number of connected components of $G$.
\end{definition}

\subsection{Shared Ancestor Tables and Vertex Depths}
Let $T=(V,E_T)$ be the rooted spanning forest, represented by
parent pointers $\Par$, with $\Par(v)=v$ at each root.
For each nonroot vertex $v$, let $e_v = (v, \Par(v))\in E_T$ denote its original
parent-edge. Shared memory stores $\Par$, $e_v$, and the tree-membership indicators
$\mathsf{isTree}[e]$.
The endpoints of each original edge occupy two consecutive
immutable input cells.
Initialize $\mathsf{isTree}[e]=0$ for all input edges, then let each
nonroot mark its original parent-edge identifier. These two phases use
$O(m+n)$ processors and constant time. Distinct nonroots have distinct
parent edges.
The depth $\depth_T(v)$ is measured from the root of the tree
containing $v$, and
$D=\max_{v\in V}\depth_T(v)$ denotes the maximum tree depth.
This maximum is used only in the analysis; the algorithm need not
compute it or split the vertex array by component.

The key idea is to construct ancestor pointers and compute vertex depths
simultaneously by pointer doubling.
The shared entry $\J[k,v]$ stores the ancestor
$2^k$ edges above $v$ or the root if the depth of $v$ is less than $2^k$. The temporary value $\Dis[v]$ stores the distance from $v$ to $\J[k,v]$. 
We initialize
\[
    \J[0,v]=\Par(v),
    \qquad
    \Dis[v]= (\Par(v)\ne v).
\]
Given level $k$, set $u=\J[k,v]$ and compute
\[
    \J[k+1,v]=\J[k,u],
    \qquad
    \Dis'[v]=\Dis[v]+\Dis[u]
\]
for every vertex in parallel.
After these computations finish, replace $\Dis$ by $\Dis'$.
Thus, at each completed level,
\[
    \J[k,v]=\Par^{(2^k)}(v),
    \qquad
    \Dis[v]=\min\{2^k,\depth_T(v)\}.
\]
The procedure continues while some $\J[k,v]$ is not a root.
At the first level $t$ where every stored ancestor is a root,
we have $\Dis[v]=\depth_T(v)$ for every vertex.
Consequently, $t=\lceil\log_2 D\rceil$ when $D \ge 1$,
and $t=0$ when $D=0$.

\paragraph{How the PRAM builds the table.}
We use a CRCW PRAM with one processor $P_v$
for each vertex $v$.
The rows of $\J$ have stride $n=|V|$ and are constructed
successively; $\Dis$ and $\Dis'$ are separate shared arrays of
length $n$.
At level $k$, processor $P_v$ reads $u=\J[k,v]$,
then reads $\J[k,u]$, $\Dis[v]$, and $\Dis[u]$.
It writes only $\J[k+1,v]$ and $\Dis'[v]$.
These writes can be performed concurrently since they write different entries.
All processors finish these computations before copying
$\Dis'[v]$ into $\Dis[v]$, ensuring that every update uses the
preceding level's distances.
A designated processor maintains the level counter and the
powers by repeated doubling.
The loop condition is evaluated in constant time: a designated
processor first clears a shared Boolean bit, and every processor
whose current ancestor is not a root then writes $1$ to that bit.
The bit is read after all tests finish.
These concurrent writes agree and are therefore legal in the
ARBITRARY CRCW model.
Each level takes $O(1)$ parallel time, giving
$O(t+1)=O(\log D)$ time and $O(n(t+1))$ work
using $O(n)$ processors.
The stored ancestor table occupies $O(n(t+1))$ words,
while depth calculation requires only $O(n)$ temporary words.

\begin{framed}
\small
\setlength{\parindent}{0pt} 

\algcaption{Ancestor Table and Vertex Depths:}\par
\label{alg:table}
\smallskip
\textbf{Input:}
Parent pointers $\Par:V\to V$ of a rooted spanning forest $T$.\par
\textbf{Output:}
Vertex depths $\depth_T(v)$, the ancestor table
$\J[k,v]=\Par^{(2^k)}(v)$ for $k\in\{0,\ldots,t\}$.\par
\smallskip
\textbf{Preprocessing ($\textsc{Preprocess}(\Par)$):}

\begin{enumerate}[
    label=\arabic*.,ref=\arabic*,
    leftmargin=1.5em,
    topsep=3pt,
    itemsep=3pt,
    parsep=0pt
]

\item
For every $v\in V$, in parallel, initialize
\[
    \J[0,v]\gets\Par(v),
    \qquad
    \Dis[v]\gets
    \begin{cases}
        0, & \Par(v)=v,\\
        1, & \Par(v)\ne v.
    \end{cases}
\]
Set $k\gets0$.

\item
\label{step:build-depth-and-ancestor}
Repeat while there exists $v\in V$ such that
$\Par(\J[k,v])\ne \J[k,v]$:
\begin{enumerate}[
    label=\alph*.,ref=\arabic{enumi}.\alph*,
    leftmargin=1.4em,
    topsep=2pt,
    itemsep=2pt,
    parsep=0pt
]

    \item
    For every $v\in V$, in parallel, set $u\gets \J[k,v]$
    and compute
    \[
        \J[k+1,v]\gets \J[k,u],
        \qquad
        \Dis'[v]\gets \Dis[v]+\Dis[u].
    \]

    \item
    After all computations in (a) finish, set
    $\Dis[v]\gets \Dis'[v]$ for every $v\in V$ in parallel.
    Set $k \gets k+1$.

\end{enumerate}

\item
Set $t\gets k$.
For every $v\in V$, in parallel, set
$\depth_T(v)\gets \Dis[v]$.
Output $\J$, $\depth_T$, and $t$.

\end{enumerate}
\end{framed}

\begin{lemma}\label{lemma:correctness-of-table-and-depth}
Algorithm~\ref{alg:table} returns the correct depths and ancestor table
for every rooted forest.
\end{lemma}
\begin{proof}
We inductively prove that at level $k$, $\J[k,v]=\Par^{(2^k)}(v)$ and
$\Dis[v]=\min\{2^k,\depth_T(v)\}$. For $k = 0$, the equations obviously hold. Suppose the equations hold before level $k+1$.
Let $u=\J[k,v]$. By the induction hypothesis, the pointer update gives
\[
\J[k+1,v]=\J[k,u]
=\Par^{(2^k)}\bigl(\Par^{(2^k)}(v)\bigr)
=\Par^{(2^{k+1})}(v).
\]
Since each root points to itself, the depth of $u$ is
$\depth_T(u)=\max\{\depth_T(v)-2^k,0\}$. The distance update therefore gives
\[
\begin{aligned}
\Dis'[v]
&=\Dis[v]+\Dis[u]\\
&=\min\{2^k,\depth_T(v)\}
+\min\{2^k,\max\{\depth_T(v)-2^k,0\}\}\\
&=\min\{2^{k+1},\depth_T(v)\}.
\end{aligned}
\]
After replacing $\Dis$ by $\Dis'$, both equations hold at level $k+1$, completing the induction.

At termination, $\J[k,v]$ is a root for every vertex $v$, so
$\depth_T(v)\le 2^k$ and hence $\Dis[v]=\depth_T(v)$.
Thus, the algorithm returns the correct ancestor table and vertex depths.
The argument applies independently to each tree, including isolated vertices.
\end{proof}

\begin{lemma}\label{lemma:complexity-of-table}
Algorithm~\ref{alg:table} takes $O(\log D)$ time with $O(n)$ processors
and $O(n\log D)$ shared words.
\end{lemma}
\begin{proof}
By Lemma~\ref{lemma:correctness-of-table-and-depth}, the algorithm constructs $t+1=O(\log D)$ levels.
Each level, including the termination test, takes constant time using one processor per vertex.
The same processors are reused at every level, giving
$O(\log D)$ total time with $O(n)$ processors.
The ancestor table stores $n$ entries per level, while the distance arrays require $O(n)$ additional words.
Thus, the total shared space is $O(n\log D)$.
\end{proof}

\subsection{Lowest Common Ancestor}

\begin{framed}
\small
\setlength{\parindent}{0pt}

\algcaption{Lowest Common Ancestor:}\par
\label{alg:lca}
\smallskip
\textbf{Input:}
Two vertices $u,v$ in the same rooted tree;
$\depth_T(u)$, $\depth_T(v)$;
$t=\lceil\log_2 D\rceil$;
shared tables $\J[k,z]=\Par^{(2^k)}(z)$ for $k\in\{0,\ldots,t\}$,
where $D\ge1$ is the maximum tree depth.\par
\textbf{Output:}
The lowest common ancestor of $u$ and $v$.\par
\smallskip
\textbf{Finding LCA ($\textsc{LCA}(u,v)$):}

\begin{enumerate}[
    label=\arabic*.,ref=\arabic*,
    leftmargin=1.5em,
    topsep=3pt,
    itemsep=3pt,
    parsep=0pt
]

\item
\label{step:lca-order}
Initialize $x\gets u$, $y\gets v$.
If $\depth_T(x)<\depth_T(y)$, exchange $x,y$.
Set $\ell\gets\depth_T(x)-\depth_T(y)$.

\item
\label{step:decrease-depth-gap}
For $k=t,t-1,\ldots,0$, if $\ell\ge 2^k$,
set
\[
    x\gets \J[k,x],
    \qquad
    \ell\gets\ell-2^{k}.
\]

\item
If $x=y$, output $x$ and terminate.

\item
\label{step:find-common-ancestor}
For $k=t,t-1,\ldots,0$, set
$a\gets \J[k,x]$ and $b\gets \J[k,y]$.
If $a\ne b$, set
\[
    x\gets a,
    \qquad
    y\gets b.
\]

\item
Output $\J[0,x]$.

\end{enumerate}
\end{framed}

\begin{lemma}\label{correctness-of-LCA}
For any vertices $u,v$ in one rooted tree, Algorithm~\ref{alg:lca}
returns their lowest common ancestor.
\end{lemma}
\begin{proof}
The exchange in \algstep{alg:lca}{step:lca-order} makes the depth
difference nonnegative. In \algstep{alg:lca}{step:decrease-depth-gap},
the binary expansion of this difference moves only the deeper vertex,
until both vertices have the same depth. This preserves their lowest
common ancestor. If they coincide, that vertex is the answer.
Otherwise, in \algstep{alg:lca}{step:find-common-ancestor}, a move is
made only when the two candidate ancestors differ. Both candidates then
remain strictly below the lowest common ancestor, and the answer is
unchanged. After processing level $k$, the distance to that ancestor is
at most $2^k$: if it were larger, the distinct $2^k$-ancestors would have
been used. At level zero the two vertices therefore have the same
parent, which is precisely the answer.
\end{proof}

\begin{lemma}\label{lemma:complexity-of-LCA}
After preprocessing, any $Q$ lowest-common-ancestor queries take
$O(\log D)$ time with $O(Q)$ processors and $O(1)$ private words per
processor.
\end{lemma}
\begin{proof}
Assign one processor to each query.
Each query makes at most two passes over the $t+1$ levels of the ancestor table, performing a constant number of operations per level.
Since concurrent reads are allowed, all $Q$ queries run in parallel and finish in $O(t+1)=O(\log D)$ time using $O(Q)$ processors.
Each processor maintains only a constant number of local variables, requiring $O(1)$ private words.
\end{proof}

\subsection{2-Edge Connectivity}
A tree edge is not a bridge exactly when it lies on a fundamental cycle formed by a non-tree edge. We mark the tree path of every non-tree edge.
The two endpoint-to-ancestor paths are processed sequentially by its
edge processor; all non-tree edges are processed in parallel.

\begin{framed}
\small
\setlength{\parindent}{0pt}
\algcaption{Finding Bridges}
\label{alg:bridge}\par
\textbf{Input:} An undirected graph $G=(V,E)$ with indexed edges.\par
\textbf{Output:} A shared array $\mathsf{bridge}[e]$ indicating whether
each input edge $e$ is a bridge.\par
\begin{enumerate}[label=\arabic*.,ref=\arabic*,leftmargin=1.5em,itemsep=3pt,parsep=0pt]
\item \label{step:bridge-forest}
Run Algorithm~\ref{alg:additive-sf} to compute a rooted spanning forest
$T$ and original parent edges $e_v$. Initialize
$\mathsf{bridge}[e]\gets0$ for every input edge.
\item \label{step:bridge-table}
Run Algorithm~\ref{alg:table} on the parent array to compute
$\J$, $\depth_T$, and $t$. For $k=0,\ldots,t$, initialize
$\Cover[k,v]\gets0$ for all vertices in parallel.
\item \label{step:bridge-paths}
For every non-tree edge $\{u,v\}$ in parallel:
\begin{enumerate}[label=\alph*.,ref=\arabic{enumi}.\alph*,leftmargin=1.4em,itemsep=2pt,parsep=0pt]
\item \label{step:bridge-lca}
Exchange $u,v$ if necessary to put the deeper endpoint first;
set $w\gets\textsc{LCA}(u,v)$ using Algorithm~\ref{alg:lca}.
\item \label{step:bridge-mark}
For each $s\in\{u,v\}$, set $x\gets s$.
For $k=t,t-1,\ldots,0$, if
$\depth_T(x)-\depth_T(w)\ge2^k$, write
$\Cover[k,x]\gets1$ and set $x\gets\J[k,x]$.
\end{enumerate}
\item \label{step:bridge-propagate}
For $k=t,t-1,\ldots,1$, process all vertices in parallel.
If $\Cover[k,v]=1$, set $a\gets\J[k-1,v]$ and write
\[
 \Cover[k-1,v]\gets1,\qquad \Cover[k-1,a]\gets1.
\]
\item \label{step:bridge-output}
For each nonroot $v$ in parallel, write
$\mathsf{bridge}[e_v]\gets[\Cover[0,v]=0]$.
Return $\mathsf{bridge}$.
\end{enumerate}
\end{framed}

Algorithm~\ref{alg:bridge} first constructs a rooted spanning forest, computes vertex depths, and builds the ancestor table. For each non-tree edge, it finds the LCA of its endpoints and splits the corresponding tree path into two paths ending at that ancestor. The \emph{marking scan} in \algstep{alg:bridge}{step:bridge-mark} visits the levels from $t$ down to $0$, marking consecutive segments whose lengths are powers of two and which together cover each path. The \emph{propagation} step in Step~\ref{step:bridge-propagate} then marks the two halves of each marked segment, proceeding down the levels until the marks reach individual tree edges. Finally, Step~\ref{step:bridge-output} reports the unmarked tree edges as bridges.

\paragraph{Shared marks and execution.}
The cell $\Cover[k,v]$ is at address
$\mathrm{base}_{\Cover}+kn+v$. It represents the $2^k$ tree edges
from $v$ up to $\J[k,v]$, excluding the edge above that ancestor.
Only segments that fit below their target ancestor are marked.
One level is initialized or propagated at a time with $O(n)$ processors;
each processor performs its two writes in separate PRAM steps.
All concurrent writes to a mark cell write the same value, one.
The level loop is sequential, so marks produced at one level are visible
at the next. A root has no output parent-edge cell.

\begin{lemma}
\label{lem:path-marking}
For a collection of $Q$ vertex-to-ancestor paths, Algorithm~\ref{alg:bridge} marks precisely their union of tree edges in
$O(\log D)$ time, with $O(Q+n)$ processors and
$O(n\log D)$ shared words.
\end{lemma}
\begin{proof}
For each input path, the marking scan divides its length into powers of two and marks the corresponding consecutive segments. These segments cover the entire path and contain no edges outside it.
During propagation, each marked segment of length $2^k$, for $k>0$, marks its two halves of length $2^{k-1}$.
The two halves contain exactly the same edges as the original segment, so propagation preserves the set of covered edges.
At level $0$, each segment consists of a single tree edge. The marked edges are therefore exactly the union of the input paths.

Marking uses one processor per path, and propagation uses one processor per vertex.
Each stage takes constant time per level, and all concurrent writes agree on the value $1$.
Since there are $t+1=O(\log D)$ levels, the total time is $O(\log D)$ with $O(Q+n)$ processors.
The shared tables store $n$ entries per level and thus use $O(n\log D)$ words.
\end{proof}

\begin{lemma}\label{lemma:correctness-of-2-edge}
Algorithm~\ref{alg:bridge} marks an input edge as a bridge if and only
if deleting that edge increases the number of connected components.
\end{lemma}
\begin{proof}
Deleting a non-tree edge leaves the spanning forest, so it is not a
bridge. Deleting a tree edge $e_v$ partitions its tree into two parts.
$e_v$ is not a bridge if and only if the tree path of some fundamental cycle formed by a non-tree edge contains it.
The LCA call in \algstep{alg:bridge}{step:bridge-lca} is valid in either
input order. The two paths in \algstep{alg:bridge}{step:bridge-mark}
partition the tree path of that non-tree edge. Lemma~\ref{lem:path-marking}
therefore makes $\Cover[0,v]=1$ exactly for non-bridge tree edges.
The output rule gives the claimed indicators. This argument applies
independently in every component, including isolated vertices.
\end{proof}

\begin{lemma}\label{lemma:complexity-of-bridge}
With good probability, Algorithm~\ref{alg:bridge} outputs all bridge
indicators in $O(\log d+\log\log n)$ time using $O(m+n)$ processors.
\end{lemma}
\begin{proof}
Theorem~\ref{thm:sf} supplies the parent array in the stated time and
gives $\log D=O(\log d+\log\log n)$. Preprocessing takes $O(\log D)$
time with $O(n)$ processors. The $m$ LCA queries and their path scans
take $O(\log D)$ time with $O(m)$ processors; propagation uses $O(n)$
processors by Lemma~\ref{lem:path-marking}. The processor pool is reused
between these stages. Output takes constant time. Conditional on a correct spanning forest, all remaining operations are deterministic. This proves
Theorem~\ref{thm:main-bridge} with the same good-probability guarantee.
\end{proof}

\section{Biconnectivity without DFS Sequence Construction}
\label{sec:bicon}
In this section, we introduce an algorithm to find biconnected blocks without constructing a DFS sequence. Following the fundamental-cycle approach of~\cite{TV85,ASZ19},
we build an auxiliary graph with one vertex for each tree edge of a
rooted spanning forest. For each fundamental cycle, we connect the
vertices representing its tree edges into a path. The connected
components of this auxiliary graph correspond exactly to the
biconnected blocks. Each non-tree edge is assigned to the component
containing the tree edges of its fundamental cycle.

We consider undirected graphs without self-loops; parallel edges are
allowed and are distinguished by their input indices. A simple cycle
has distinct edges and no repeated vertex except its start and end, and two parallel edges
may form a cycle of length two. We assume $n\ge2$, since smaller graphs
have no edges to label.

\begin{definition}[Biconnected blocks]
\label{def:blocks}
For edges $e,f\in E$, write $e\sim f$ if $e=f$ or there is a simple
cycle containing both edges. The equivalence classes of this relation
are the biconnected blocks. A bridge forms a singleton block. Blocks
partition the edge set and may share cut vertices; isolated vertices
have no edges to label.
\end{definition}

To see why this relation is transitive, consider two cycles sharing an
edge. After deleting any vertex, the remaining part of each cycle is
connected, and at least one endpoint of their shared edge still joins
the two parts. Their union therefore has no cut vertex. Subdivide any
two distinct edges of this union. The resulting graph still has no cut
vertex, so Menger's theorem gives two internally vertex-disjoint paths
between the new subdivision vertices. These paths form a simple cycle
through both original edges. This proves transitivity; reflexivity and
symmetry follow directly from the definition.

\begin{definition}[Bi-diameter]
\label{def:bi-diameter}
The bi-diameter $d'$ is the maximum, over all pairs of distinct vertices
lying on a common simple cycle, of the length of a shortest simple
cycle containing both vertices. If $G$ has no cycle, set $d'=1$.
\end{definition}

Let $T=(V,E_T)$ be a rooted spanning forest of $G$, and let
$D=\max_{v\in V}\depth_T(v)$. As in Section~\ref{sec:bridge}, each root
points to itself. Let $V'$ be the set of all nonroot
vertices of $T$. For a non-tree edge $e$, write $F_e$ for its fundamental
cycle with respect to $T$.

We construct $G'=(V',E')$ as follows. Each nonroot vertex $v$ represents its original parent edge $e_v = (v, \Par(v)$. For each non-tree edge
$e=\{u,v\}$, let $w=\LCA_T(u,v)$. If $w\notin\{u,v\}$, add
$\{u,v\}$ to $E'$. Along each endpoint-to-$w$ path, also connect the
vertices representing consecutive tree edges: add $\{x,\Par(x)\}$
whenever $\Par(x)$ lies strictly below $w$ on that path. Together,
these connections form a path through the representatives of the tree
edges of $F_e$. Algorithm~\ref{alg:block} constructs these connections
using the ancestor table and shared path marks from
Section~\ref{sec:bridge}.

\begin{framed}
\small
\setlength{\parindent}{0pt}
\algcaption{Finding Biconnected Blocks:}\par
\label{alg:block}
\smallskip
\textbf{Input:}
An undirected graph $G=(V,E)$ with indexed edges; a rooted spanning
forest $T=(V,E_T)$ with parent pointers and original parent-edge
identifiers $e_v$; and the tables $\J,\depth_T$ with level bound $t$
from Algorithm~\ref{alg:table}.\par
\textbf{Output:}
A labeling $\mathsf{block}:E\to V$ such that two edges have the same
label if and only if they belong to the same block.\par
\smallskip
\begin{enumerate}[
    label=\arabic*.,ref=\arabic*,leftmargin=1.5em,
    topsep=3pt,itemsep=3pt,parsep=0pt
]
\item
\label{step:block-initialize}
Initialize $G'=(V',E')$, where
$V'=\{v\in V:\Par(v)\ne v\}$ and $E'=\emptyset$.
For $k=0,\ldots,t$, initialize $\Cover[k,v]\gets0$ for all $v\in V$
in parallel, one level at a time.

\item
\label{step:block-paths}
For every non-tree edge $e=\{u,v\}\in E\setminus E_T$, in parallel:
\begin{enumerate}[
    label=\alph*.,ref=\arabic{enumi}.\alph*,leftmargin=1.4em,
    topsep=2pt,itemsep=2pt,parsep=0pt
]
\item
\label{step:block-lca}
Exchange $u,v$ if necessary to put the deeper endpoint first; set $w\gets\textsc{LCA}(u,v)$ using Algorithm~\ref{alg:lca}.

\item
\label{step:block-truncate}
For each $s\in\{u,v\}$ in turn, set $x\gets s$.
For $k=t,t-1,\ldots,0$, if
$\depth_T(x)-\depth_T(w)>2^k$, set
\[
    \Cover[k,x]\gets1,
    \qquad x\gets\J[k,x].
\]

\item
\label{step:block-cross}
If $w\notin\{u,v\}$, add $\{u,v\}$ to $E'$.
\end{enumerate}

\item
\label{step:block-propagate}
For $k=t,t-1,\ldots,1$, process every $v\in V$ in parallel.
If $\Cover[k,v]=1$, set $a\gets\J[k-1,v]$ and write
\[
    \Cover[k-1,v]\gets1,
    \qquad \Cover[k-1,a]\gets1.
\]
Perform the two writes in separate PRAM steps, and finish each level
before proceeding to the next.

\item
\label{step:block-vertical}
For every $v\in V'$ in parallel, if $\Cover[0,v]=1$, add
$\{v,\Par(v)\}$ to $E'$.

\item
\label{step:block-cc}
Compute the connected components of $G'$ using
Theorem~\ref{theorem:connected-component} and the edge-array
representation described below. Let $\mathsf{col}'(v)$ be a vertex
label identifying the component containing $v\in V'$.

\item
\label{step:block-output}
For every edge $e=\{u,v\}\in E$ in parallel, set
\[
    \mathsf{block}[e]\gets
    \begin{cases}
        \mathsf{col}'(u), & \depth_T(u)\ge\depth_T(v),\\
        \mathsf{col}'(v), & \text{otherwise}.
    \end{cases}
\]
Return $\mathsf{block}$.
\end{enumerate}
\end{framed}

The \emph{marking scan} in
\algstep{alg:block}{step:block-truncate} visits the ancestor-table
levels from $t$ down to $0$. On a branch of length $h$, the strict
inequality makes it mark exactly the lowest $\max\{h-1,0\}$ tree
edges, leaving out the edge incident to the LCA. The \emph{propagation}
in \algstep{alg:block}{step:block-propagate} passes these marks to
individual tree edges, as in Lemma~\ref{lem:path-marking}.
For a nonroot vertex $v$, $\Cover[0,v]=1$ exactly when $\Par(v)$ is
also nonroot and $e_v,e_{\Par(v)}$ are consecutive edges on the same
branch of some fundamental cycle. Thus both endpoints of every edge
added in \algstep{alg:block}{step:block-vertical} belong to $V'$.

\begin{lemma}\label{lemma:color-to-connected}
If distinct vertices $u,v\in V'$ are in the same connected component
of $G'$, then there is a simple cycle containing both $e_u$ and $e_v$.
\end{lemma}
\begin{proof}
First suppose $\{u,v\}\in E'$. If this edge is added in
\algstep{alg:block}{step:block-cross}, its endpoints are both strictly
below their LCA. Hence $e_u$ and $e_v$ lie on the fundamental cycle of
the non-tree edge that adds it. Otherwise, it is added in
\algstep{alg:block}{step:block-vertical}. Suppose $u=\Par(v)$.
By the marking rule and Lemma~\ref{lem:path-marking}, $e_v$ and $e_u$
are consecutive tree edges on a fundamental cycle. Thus, in either
case, the two represented edges belong to the same block.

Now consider any path from $u$ to $v$ in $G'$. The argument above
applies to each pair of consecutive vertices on the path. By
transitivity of the common-cycle relation, all their represented tree
edges belong to the same block. In particular, there is a simple cycle
containing both $e_u$ and $e_v$.
\end{proof}

\begin{lemma}\label{lem:fundamental-intersection}
Let $C$ be a simple cycle in $G$, and let $X=E(C)\setminus E_T$
be its set of non-tree edges with respect to the spanning forest $T$.
For each $e\in X$, let $F_e$ be the fundamental cycle formed by $e$
and the tree path between its endpoints. Then:
\begin{enumerate}[label=(\roman*),leftmargin=*,nosep]
\item Every tree edge of $C$ belongs to at least one $F_e$ with $e\in X$.
\item Any two cycles in $\{F_e \ | \ e \in X\}$ can be linked by a sequence
of cycles from this same collection, with consecutive cycles sharing
a tree edge.
\end{enumerate}
\end{lemma}

\begin{proof}
We first show that
\[
E(C)=\bigoplus_{e\in X}E(F_e),
\]
where $\oplus$ denotes symmetric difference. Each $F_e$ contains
exactly one non-tree edge, namely $e$, so both sides contain precisely
the non-tree edges in $X$. Their symmetric difference therefore
contains only tree edges. It also has even degree at every vertex,
being a symmetric difference of cycles. A nonempty set of tree edges
forms a forest with a vertex of degree one, so this symmetric
difference must be empty. This proves the equation and hence (i).

Suppose (ii) fails, and choose $e_0,f_0\in X$ whose fundamental
cycles cannot be linked by such a sequence.
Let $Y\subseteq X$ consist of the edges whose fundamental cycles
can be linked to $F_{e_0}$ as in (ii), and let $Z=X\setminus Y$.
Both sets are nonempty, since $e_0\in Y$ and $f_0\in Z$.
No $F_e$ with $e\in Y$ shares a tree edge with an $F_f$ with $f\in Z$:
such a shared edge would extend a sequence to $F_f$, putting $f$ in $Y$.

Define the edge sets
\[
D_Y=\bigoplus_{e\in Y}E(F_e),
\qquad
D_Z=\bigoplus_{e\in Z}E(F_e).
\]
Both sets are nonempty, as they contain the non-tree edges in $Y$ and
$Z$, respectively. They are disjoint: the two collections of cycles
share no tree edges, and their non-tree edges are distinct.
The equation above therefore gives $D_Y\cup D_Z=E(C)$.
Moreover, each set defines a subgraph in which every vertex has even
degree, since it is a symmetric difference of cycles.
This is impossible: a proper nonempty subset of the edges of a simple
cycle has a vertex of degree one. This contradiction proves (ii).
\end{proof}

\begin{lemma}\label{lem:short-edge-cycle}
Any two distinct edges in the same block lie on a simple cycle of
length at most $3d'$.
\end{lemma}
\begin{proof}
Let $e=\{a,b\}$ and $f=\{c,d\}$ be the two edges, naming their
endpoints so that $a\ne c$. Since $e$ and $f$ belong to the same
block, they lie on a common simple cycle. By the definition of $d'$,
any two distinct endpoints of these edges therefore lie on a simple
cycle of length at most $d'$. In particular, choose such a cycle $C_0$
through $a$ and $c$.

First, we construct a simple cycle $C_1$ containing $e$ and $c$, with
length at most $2d'$. If $e\in E(C_0)$, take $C_1=C_0$.
If $e\notin E(C_0)$ but $b\in V(C_0)$, the two $b$--$a$ paths along
$C_0$ together cover $C_0$, so at least one contains $c$. That path
together with $e$ forms the required cycle $C_1$, of length at most
$|E(C_0)|$.

If $b\notin V(C_0)$, choose a simple cycle $Q_1$ through $b,c$ of
length at most $d'$. Its two $b$--$c$ paths are internally
vertex-disjoint, so at least one avoids $a$. Follow that path from
$b$ to its first vertex $x$ on $C_0$, and denote this initial part
by $P_1$. Then $x\ne a$, the internal vertices of $P_1$ lie outside
$C_0$, and $|E(P_1)|\le d'-1$. Choose an $x$--$a$ path along $C_0$
that contains $c$. This path, $P_1$, and $e$ form a simple cycle
$C_1$: the path from $a$ through $e$ and $P_1$ meets $C_0$ only at
$a,x$. The cycle contains both $e$ and $c$, and
\[
    |E(C_1)|\le 1+(d'-1)+|E(C_0)|\le 2d'.
\]

Next, we construct a simple cycle $C_2$ containing both $e$ and $f$.
If $f\in E(C_1)$, take $C_2=C_1$.
If $f\notin E(C_1)$ but $d\in V(C_1)$, choose the $d$--$c$ path
along $C_1$ that contains $e$. Together with $f$, this gives a simple
cycle $C_2$ of length at most $|E(C_1)|$.

If $d\notin V(C_1)$, choose a simple cycle $Q_2$ through $d,a$ of
length at most $d'$. Its two $d$--$a$ paths are internally
vertex-disjoint, and $c$ is distinct from both endpoints, so at least
one avoids $c$. Follow that path from $d$ to its first vertex
$y$ on $C_1$, obtaining a path $P_2$ of length at most $d'-1$.
Then $y\ne c$, and the internal vertices of $P_2$ lie outside $C_1$.
Of the two $y$--$c$ paths along $C_1$, choose the one containing $e$.
Together with $P_2$ and $f$, it forms a simple cycle $C_2$ by the
same reasoning as above. This cycle contains both prescribed edges,
and
\[
    |E(C_2)|\le 1+(d'-1)+|E(C_1)|\le 3d'.
\]
Thus the required cycle exists in every case.
\end{proof}

\begin{lemma}\label{lemma:correctness-of-biconnectivity}
Let $G=(V,E)$ be an undirected graph and let $T=(V,E_T)$ be a rooted
spanning forest of maximum depth $D$. For distinct edges $e_1,e_2$,
Algorithm~\ref{alg:block} satisfies
$\mathsf{block}[e_1]=\mathsf{block}[e_2]$ if and only if there is a
simple cycle containing both edges. Furthermore, every connected
component of $G'$ has diameter at most $6Dd'$.
\end{lemma}
\begin{proof}
First we show that edges with the same label belong to the same block.
For an original edge $e$, let $e^*$ be the parent edge of the endpoint
chosen in \algstep{alg:block}{step:block-output}. That endpoint is
always a nonroot vertex. If $e$ is a tree edge, then $e^*=e$.
Otherwise, $e$ and $e^*$ lie on the fundamental cycle $F_e$.
Thus $e$ and $e^*$ belong to the same block and receive the same label.
If two original edges have the same label, the auxiliary vertices
representing their chosen parent edges are equal or connected in
$G'$. Lemma~\ref{lemma:color-to-connected} shows that these parent
edges belong to the same block, and hence so do the original edges.

Next we show that all edges on a simple cycle receive the same label.
For a non-tree edge $e=\{u,v\}$, let $w=\LCA_T(u,v)$.
The edges added for $F_e$ connect consecutive tree-edge representatives
along each endpoint-to-$w$ path. If both branches are nonempty,
$\{u,v\}$ joins their lower endpoints. These connections form a path
$P_e$ containing every tree-edge representative of $F_e$.
If $F_e$ has just one tree edge, $P_e$ is a single vertex.
In all cases, $P_e$ has at most $2D$ edges.
The output rule also assigns $e$ the label of a vertex on $P_e$.

Let $C$ be any simple cycle. By
Lemma~\ref{lem:fundamental-intersection}, the union of the paths
$P_e$, for $e\in E(C)\setminus E_T$, is connected: a tree edge
shared by two fundamental cycles gives a vertex shared by their
auxiliary paths. This union contains the representatives of all tree
edges of $C$, and each non-tree edge of $C$ receives a label from one
of these paths. Hence all edges of $C$ receive the same label.
Together with the first part, this proves the required equivalence.
It also shows that each bridge receives a label shared by no other
edge.

Finally, take distinct vertices $x,y$ in one component of $G'$.
By Lemma~\ref{lemma:color-to-connected}, their tree edges belong to
the same block. Lemma~\ref{lem:short-edge-cycle} gives a simple cycle
$C$ through both edges with $|E(C)|\le3d'$.
The connected union of its auxiliary paths contains $x,y$ and has at
most
\[
    \sum_{e\in E(C)\setminus E_T}|E(P_e)|
    \le 2D\,|E(C)\setminus E_T|
    \le 6Dd'
\]
edges. It therefore contains an $x$--$y$ path of length at most
$6Dd'$. Singleton components have diameter zero.
The argument applies independently to every component of $G$;
isolated vertices have no edges to label.
\end{proof}

\paragraph{Build auxiliary graph on PRAM.}
We implement the additions to $E'$ using an array $A$ indexed by the
original edges. Initially, $A[e]$ contains a self-loop at one endpoint
of $e$. An edge added in \algstep{alg:block}{step:block-cross} is
written to $A[e]$, and an edge added in
\algstep{alg:block}{step:block-vertical} is written to $A[e_v]$.
These entries have distinct writers: the first step uses non-tree
edge indices, and the second uses tree-edge indices.
Thus the array can be constructed without counting the auxiliary
edges or moving them to consecutive positions.

All original vertex indices are retained, with roots as isolated
vertices. Self-loops and these isolated vertices change neither
connectivity on $V'$ nor the maximum component diameter.

\begin{theorem}[\cite{LTZ20}]\label{theorem:connected-component}
There is an ARBITRARY CRCW PRAM algorithm that computes the connected
components of an undirected graph with $n$ vertices, $m$ edges, and
maximum component diameter $d$, using $O(m+n)$ processors.
With probability at least $1-1/\poly((m+n)\log n/n)$, it runs in
$O(\log d+\log\log n)$ time.
\end{theorem}

\begin{lemma}\label{lemma:complexity-of-bicomponent}
Given a rooted spanning forest of maximum depth $D$,
Algorithm~\ref{alg:block}, including ancestor-table preprocessing,
takes
\[
    O\bigl(\log D+\log d'+\log\log n\bigr)
\]
time with good probability using $O(m+n)$ processors.
Including forest construction, the total time is
\[
    O\bigl(\log d+\log d'+\log\log n\bigr),
\]
where $d$ is the maximum component diameter of $G$.
\end{lemma}
\begin{proof}
Tree membership can be marked from the original parent-edge identifiers
in constant time with $O(m+n)$ processors, as in
Section~\ref{sec:bridge}. The ancestor table, LCA queries, marking
scans, and propagation take
$O(\log D)$ time with $O(m+n)$ processors by the results in
Section~\ref{sec:bridge}. One processor handles each non-tree edge
and processes its two branches in turn. Initialization and propagation
use one processor per vertex, reused at each level. All concurrent
writes to a mark entry write $1$.
Constructing the auxiliary edge array and assigning the final edge
labels each take constant time with $O(m+n)$ processors.

The graph $ G'$ has $n$ vertices, $m$ edge entries,
and maximum component diameter at most $6Dd'$ by
Lemma~\ref{lemma:correctness-of-biconnectivity}.
Theorem~\ref{theorem:connected-component} therefore computes its
components in
$O(\log D+\log d'+\log\log n)$ time with $O(m+n)$
processors and failure probability
$1/\poly((m+n)\log n/n)$.
This proves the first bound.

By Theorem~\ref{thm:sf}, forest construction takes
$O(\log d+\log\log n)$ time with good probability and gives
$\log D = O(\log d+\log\log n)$.
Substituting this bound proves the claimed total running time.
Use fresh random bits for the auxiliary connected-components call.
Conditioned on any successful forest construction, its failure bound
holds for the resulting auxiliary graph. Averaging over the possible
forests and taking a union bound with the failure of forest
construction preserves good probability. All stages reuse the same
$O(m+n)$ processors. This establishes Theorem~\ref{thm:main-block}.
\end{proof}

\section{Lower Bound}
In this section, we establish deterministic parallel time lower bounds for computing a DFS sequence, a weighted spanning forest, and an open ear decomposition on an ARBITRARY CRCW PRAM with polynomially many processors. Our starting point is the parity problem: given bits $x_1,\ldots,x_k\in \{0,1\}$, compute $\bigl(\sum_{i=1}^{k}x_i\bigr)\bmod 2$. We first recall the known lower bound for parity and then present a reduction from parity to each of the three graph problems. These reductions preserve the asymptotic parallel time and use polynomially many processors, thereby transferring the parity lower bound to all three problems.

\begin{definition}[PARITY]
For a positive integer \(k\) and \(x\in\{0,1\}^k\), let
\(\operatorname{PARITY}(x)=(\sum_{i=1}^k x_i)\bmod2\).
\end{definition}

\begin{theorem}[\cite{BH89}]\label{thm:lower-bound-of-parity}
    Let $x \in \{0,1\}^k$ be input bits initially stored in separate shared-memory cells. Any deterministic PRIORITY CRCW PRAM algorithm using $k^{O(1)}$ processors that computes
    \[
        \operatorname{PARITY}(x)
        = \left(\sum_{i=1}^{k} x_i\right) \bmod 2
    \]
    correctly on every input requires
    \[
        \Omega\!\left(\frac{\log k}{\log\log k}\right)
    \]
    parallel time in the worst case. Consequently, the same lower bound holds for the ARBITRARY CRCW PRAM.
\end{theorem}

\subsection{DFS Sequence}
A traversal can be represented implicitly by successor pointers
or explicitly by one shared array cell for every visit. These
representations have different computational requirements. An explicit
position records the sizes of all traversal intervals preceding it.
We make this obstruction precise by encoding PARITY in a subtree
interval of a depth-two tree. The reduction works for every child order,
so choosing an easier order does not remove the obstruction.

\begin{definition}[DFS sequence]
\label{def:dfs-sequence}
Let \(T=(V,E_T)\) be a rooted tree with
\(V=\{0,\ldots,n-1\}\) and specified root \(r\).
Its parent mapping \(p:V\to V\) satisfies \(p(r)=r\);
for \(v\ne r\), \(p(v)\) is the neighbor of \(v\) on the unique
path from \(v\) to \(r\). Define
\[
 \mathcal C_p(v)=\{u\in V\setminus\{v\}:p(u)=v\}.
\]
A child-order family \(\sigma\) specifies, for every vertex \(v\),
a permutation \(\sigma_v=(v_1,\ldots,v_t)\) of
\(\mathcal C_p(v)\), where \(t=|\mathcal C_p(v)|\).
The permutation is empty for a leaf.
Writing \(\circ\) for sequence concatenation, define
\[
 \mathcal W_\sigma(v)=
 \begin{cases}
  (v),&t=0,\\
  (v)\circ\mathcal W_\sigma(v_1)\circ(v)\circ\cdots
      \circ\mathcal W_\sigma(v_t)\circ(v),&t\ge1.
 \end{cases}
\]
The DFS sequence of \(T\) is \(\mathcal W_\sigma(r)\): it records
the initial visit to the root and the vertex reached after every
edge traversal, including returns from a child to its parent.

The computational input consists only of \(n\), the specified root
\(r\), and the shared parent array \(p[0\,..\,n-1]\), promised to
represent a rooted tree. The algorithm may choose any family
\(\sigma\). It must write a shared contiguous array
\(S[0\,..\,2n-2]\) satisfying
\[
 S[j]=\mathcal W_\sigma(r)[j]\qquad(0\le j\le2n-2)
\]
for one such family, with zero-based sequence indices.
Position \(j\) occupies address \(\mathrm{base}_S+j\);
the output base is specified or returned in one shared word.
\end{definition}

\paragraph{Interval property of the recursive definition.}
Let \(n_v\) count the vertices of \(v\)'s subtree, including \(v\).
Induction gives
\[
 |\mathcal W_\sigma(v)|
 =1+\sum_{u\in\mathcal C_p(v)}
       \bigl(|\mathcal W_\sigma(u)|+1\bigr)
 =2n_v-1.
\]
Each child sequence is inserted as one contiguous subsequence.
Because every nonroot vertex has exactly one parent, expansion of
\(\mathcal W_\sigma(r)\) invokes the recursion at \(v\) exactly once.
Only that invocation directly emits copies of vertex \(v\);
the invocations at other vertices directly emit their own identifiers.
Thus all occurrences of \(v\) lie in its unique recursively generated
block \(\mathcal W_\sigma(v)\), whose first and last entries are \(v\).
Consequently the first and last occurrences
\[
 \operatorname{first}_S(v)=\min\{j:S[j]=v\},\qquad
 \operatorname{last}_S(v)=\max\{j:S[j]=v\}
\]
are exactly the boundaries of \(\mathcal W_\sigma(v)\) in the
full output. Consequently,
\begin{equation}
 \operatorname{last}_S(v)-\operatorname{first}_S(v)=2(n_v-1).
 \label{eq:dfs-interval}
\end{equation}
For a leaf these boundaries coincide.
The full array has exactly \(2n-1\) entries, rather than an
unspecified \(O(n)\) number of visits.
The input supplies no degrees, compact child lists, child ranks,
subtree sizes, traversal positions, or Euler successor list.
Such data, if needed, must be computed by the algorithm.

\paragraph{Processor convention.}
For a tree, \(m=n-1\), so our \(O(m+n)\) budget is \(O(n)\).
Fix an available pool of \(P(n)=\lceil\gamma n\rceil\) processors for
a constant \(\gamma\ge1\); unused processors may remain idle.
All stages of the reduction execute sequentially on this same
pool. No separate pool is allocated for construction or extraction.

\begin{theorem}[PARITY reduction]
\label{thm:parity}
Suppose an ARBITRARY CRCW PRAM algorithm computes a
DFS sequence as in Definition~\ref{def:dfs-sequence} in \(T(n)\)
time on \(O(n)\) processors.
For \(n=k+3\), PARITY on \(k\) input bits can be computed in
\(T(n)+O(1)\) time with the same asymptotic processor bound.
If the traversal algorithm has success probability \(\rho(n)\),
the reduction has success probability at least \(\rho(n)\).
\end{theorem}
\begin{proof}
Create vertices \(r,a,c,b_1,\ldots,b_k\), with identifiers
\(r=0,a=1,c=2,b_i=i+2\), and parents
\[
 p(r)=r,\qquad p(a)=p(c)=r,\qquad
 p(b_i)=
 \begin{cases}a,&x_i=1,\\c,&x_i=0.\end{cases}
\]
The tree has \(n=k+3\) vertices, depth two, and
\(n_a=1+\sum_i x_i\).

During construction, processor \(P_{i-1}\) reads \(x_i\) from
\(\mathrm{base}_x+i-1\) and writes the one shared parent cell for
\(b_i\). In a separate constant-size phase, three processors
from the same pool write the fixed parents of \(r,a,c\).
Thus construction takes \(O(1)\) time on the available pool.

Next reuse the pool to run the assumed DFS algorithm.
After it finishes, reuse the pool for extraction.
There are \(2n-3\) interior output positions, which need not be
examined simultaneously. In batch one, processor \(P_i\),
\(0\le i<n\), examines position \(j=i+1\). In batch two it examines
\(j=i+n+1\). In either batch it is idle when \(j>2n-3\).
The two batches cover every \(1\le j\le2n-3\) exactly once,
using only the first \(n\) processors of the same pool.

For an assigned position, the processor reads
\(S[j-1],S[j],S[j+1]\) in three PRAM read steps.
One existing processor first initializes two shared scalar cells
to \(-1\).
A match \(S[j]=a,S[j-1]=r\) writes \(j\) to \(j_{\rm in}\);
a match \(S[j]=a,S[j+1]=r\) writes \(j\) to \(j_{\rm out}\).
For a valid sequence, each match is unique.
The first occurrence of \(a\) is preceded by \(r\), and the last
is followed by \(r\); all other neighbors of an occurrence of
\(a\) within its subtree are its children.
When \(a\) is a leaf, both matches occur at the same position;
the processor writes the two distinct scalar cells in separate steps.
After both batches, an existing processor reads the results.

By Equation~\eqref{eq:dfs-interval},
\[
 j_{\rm out}-j_{\rm in}=2\sum_{i=1}^k x_i.
\]
Return \(((j_{\rm out}-j_{\rm in})/2)\bmod2\).
The child order and the relative order of \(a,c\) do not affect
this equation. On an incorrect traversal output, multiple matches
remain legal under arbitrary write resolution. If a scalar is
unset, the bounds are invalid, or the difference is negative or odd,
return the bit \(0\). Extraction therefore always takes \(O(1)\)
time without validating the entire traversal.

Every correct DFS output gives correct PARITY, so success probability
is at least \(\rho(k+3)\); it need not be exactly equal, since an
incorrect traversal can accidentally give correct parity.
Construction, DFS computation, and extraction run successively.
Their processor requirements are bounded by the same \(P(n)\),
rather than added, and their times sum to \(T(k+3)+O(1)\).
\end{proof}

\begin{corollary}[lower bound for DFS Sequence]
\label{cor:dfs}
With \(O(n)\) processors, deterministic ARBITRARY CRCW PRAM DFS sequence construction requires
\(\Omega(\log n/\log\log n)\) worst-case time.
\end{corollary}
\begin{proof}
Apply Theorem~\ref{thm:parity} to the lower bound of Theorem \ref{thm:lower-bound-of-parity}. The reduction reuses a linear
pool, which is polynomially bounded in \(k=n-3\).
The cited lower bound holds for the stronger PRIORITY CRCW model
and therefore applies to our ARBITRARY CRCW model as well.
\end{proof}

It is not a matching \(O(\log n/\log\log n)\) upper bound.
It does not apply directly to an implicit successor representation,
to preprocessed input containing ranks, or to MPC.
An ordered-label representation need not expose the integer
positions used by the reduction. The probability-preserving statement
of Theorem~\ref{thm:parity} also permits applying a randomized PARITY
lower bound, such as \cite{KSParity}, whenever its model and error
hypotheses match the proposed traversal algorithm.
If the input already contains compact child lists or vertex degrees,
the bit-to-input construction must be reconsidered.

The same issue constrains exact-size bridge output.
For each input bit \(x_i\), introduce vertices \(a_i,b_i,c_i\)
and a common root \(r\). Add
\(\{r,a_i\},\{a_i,b_i\},\{b_i,c_i\}\); the fourth edge is
\(\{c_i,r\}\) if \(x_i=0\), and \(\{c_i,a_i\}\) if \(x_i=1\).
This gives \(n=3k+1,m=4k,d\le4\), and exactly \(\sum_i x_i\)
bridges. Thus returning the exact bridge count in a single word
also yields PARITY. This is why the main bridge theorem specifies
indexed indicators, rather than a packed array with an exact length.
Approximate compaction with holes is a different output contract.

\subsection{Weighted Spanning Forest}
We next consider minimum spanning forests and bottleneck spanning
forests. A reduction from PARITY gives a
\(\Omega(\log N/\log\log N)\) time lower bound.
The constructed graph is connected and has diameter two. It has only
two edge weights, and the input records are sorted by weight.
Thus the lower bound holds even when no sorting is needed.

\begin{definition}[Weighted spanning-forest guarantees]
For a connected weighted graph \(G\), let
\[
 W^*(G)=\min_{T\text{ spanning tree}}\sum_{e\in T}w(e),
 \qquad
 B^*(G)=\min_{T\text{ spanning tree}}\max_{e\in T}w(e).
\]
An \(\alpha\)-approximate MSF algorithm returns a spanning tree \(T\)
with \(W(T)\le\alpha W^*(G)\). For the bottleneck objective, it
returns \(T\) with \(B(T)\le\alpha B^*(G)\).
For a disconnected graph, apply these definitions to each component.
The reduction uses a connected graph.
\end{definition}

\begin{framed}
\small
\setlength{\parindent}{0pt}
\algcaption{The parity gadget for weighted spanning forests:}
\label{alg:weighted-parity-gadget}\par
\smallskip
\textbf{Input:}
A bit vector \(x[1..k]\) with \(k\ge2\), an integer \(B>1\), and the shared bases
\(\mathsf{base}_x,\mathsf{base}_E\).\par
\textbf{Output:}
A connected simple weighted graph \(G_x\) with \(N=2k+3\) vertices.\par
\smallskip
\begin{enumerate}[label=\arabic*.,leftmargin=1.5em,topsep=3pt,itemsep=3pt,parsep=0pt]
\item
Create vertices \(v_i^b\) for \(0\le i\le k\), \(b\in\{0,1\}\),
and one vertex \(z\). The identifier of \(v_i^b\) is \(2i+b\), and
\(z\) has identifier \(2k+2\).
\item
For every \(i\in\{1,\ldots,k\}\) and \(b\in\{0,1\}\), write the
unit-weight edge
\[
 \{v_{i-1}^b,v_i^{\,b\oplus x_i}\}.
\]
Also write \(\{v_0^0,v_k^0\}\) and \(\{z,v_0^0\}\), both of weight
one. These \(2k+2\) light edges occupy the first input records.
\item
For every \(v_i^b\ne v_0^0\), write the edge
\(\{z,v_i^b\}\) of weight \(B\). These \(2k+1\) heavy edges occupy the
remaining records.
\end{enumerate}
\end{framed}

Assign one processor to each bit-dependent edge.
Processor \(P_{2(i-1)+b}\) writes its light-edge record at the
corresponding input address. The other records have fixed indices and
are written by an \(O(N)\)-processor pool. No prefix sum or compaction
is needed. All light records precede all heavy records, so the input
is sorted by weight. The construction has \(M=4k+3=2N-3\) records
and takes \(O(1)\) time with \(O(N)\) processors.

\begin{lemma}[Structure of the parity gadget]
\label{lem:weighted-parity-gadget}
Let \(p=x_1\oplus\cdots\oplus x_k\). If \(p=1\), the light edges of
Algorithm~\ref{alg:weighted-parity-gadget} form a spanning tree and
\[
 W^*(G_x)=N-1,\qquad B^*(G_x)=1.
\]
If \(p=0\), the light graph has two components and one cycle.
Every spanning tree then contains a heavy edge, and
\[
 W^*(G_x)=N-2+B,\qquad B^*(G_x)=B.
\]
\end{lemma}
\begin{proof}
The \(2k\) layer edges form two vertex-disjoint paths.
The path starting at \(v_0^0\) ends at \(v_k^p\); the other starts
at \(v_0^1\) and ends at \(v_k^{1\oplus p}\).

If \(p=1\), the edge \(\{v_0^0,v_k^0\}\) joins the two paths.
The edge \(\{z,v_0^0\}\) attaches \(z\). The \(N-1\) light edges
therefore form a spanning tree, giving the first pair of equalities.

If \(p=0\), the closing edge creates a cycle on the first path.
The second path remains a separate component. Only heavy edges join
the two components, so every spanning tree contains one.
Remove a light edge from the cycle and add a heavy edge between the
components. This gives a spanning tree with \(N-2\) unit edges and
one edge of weight \(B\), proving the second pair of equalities.
\end{proof}

\begin{theorem}[Lower bounds for weighted spanning forests]
\label{thm:mst-lower}
Fix a constant \(\alpha\ge1\). On ARBITRARY CRCW PRAMs with
\(O(N)\) processors and \(O(1)\) private words per processor,
every deterministic algorithm for any of the following problems requires
\[
\Omega\!\left(\frac{\log N}{\log\log N}\right)
\]
time in the worst case:
\begin{enumerate}[label=(\roman*),leftmargin=1.5em,itemsep=2pt,parsep=0pt]
\item exact MST, and therefore exact MSF;
\item \(\alpha\)-approximate MSF, including fixed
\((1+\epsilon)\)-approximation for every constant \(\epsilon>0\);
\item \(\alpha\)-approximate bottleneck MSF.
\end{enumerate}
The bound holds even for connected diameter-two graphs with two weights
and an input array whose light records precede its heavy records.
\end{theorem}
\begin{proof}
We reduce PARITY to each problem and apply the deterministic bound
of Theorem~\ref{thm:lower-bound-of-parity}. Given \(x\in\{0,1\}^k\), construct \(G_x\) using
Algorithm~\ref{alg:weighted-parity-gadget}. Construction, the assumed
forest algorithm, and decoding reuse the same \(O(N)\) processors.

For total weight, choose \(B>\alpha(N-1)\). If \(p=1\), the optimum
has weight \(N-1\), so an \(\alpha\)-approximate output cannot contain
a heavy edge. If \(p=0\), every spanning tree contains a heavy edge.
For the bottleneck objective, choose \(B>\alpha\).
When \(p=1\), an \(\alpha\)-approximate output has bottleneck at most
\(\alpha\), so it contains no heavy edge. When \(p=0\), every
spanning tree contains a heavy edge.

To decode, initialize one shared flag to zero. Assign one processor
to each output record. Any processor that finds a selected heavy edge
writes \(1\) to the flag. These concurrent writes are identical.
The flag equals \(1-p\), and decoding takes \(O(1)\) time.
Construction also takes \(O(1)\) time. An
\(o(\log k/\log\log k)\)-time forest algorithm would therefore
contradict the PARITY lower bound. Since \(N=\Theta(k)\), the claimed
bound follows. Taking \(\alpha=1\) gives exact MST, and taking
\(\alpha=1+\epsilon\) gives the approximation bound.
\end{proof}

\subsection{Open Ear Decomposition}
We reduce parity to open ear decomposition by constructing a graph consisting of four internally vertex-disjoint paths with common endpoints $s$ and $t$. The input bits determine how the vertices on consecutive layers are connected. Their parity determines which vertex next to $t$ is reached by a path starting at a specified vertex next to $s$. Every open ear decomposition identifies at least one of these paths as an individual ear. We can therefore recover the parity by inspecting the ear labels of the edges incident to $s$ and $t$.

\begin{definition}[Open ear decomposition]
\label{def:open-ear-decomposition}
Let $G=(V,E)$ be a simple undirected graph. An \emph{open ear decomposition} of $G$ is a sequence $E_1,\ldots,E_q$ of nonempty subsets of $E$ satisfying
the following conditions:
\begin{enumerate}[label=(\arabic*),leftmargin=*]
    \item
    The sets $E_1,\ldots,E_q$ partition $E$, and their incident vertices together cover $V$.
    \item
    The edges in $E_1$ form a simple cycle.
    \item
    For each $i\geq 2$, the edges in $E_i$ form a simple path with two distinct endpoints. Both endpoints belong to
    \[
        \bigcup_{j<i} V(E_j),
    \]
    where $V(E_j)$ denotes the vertices incident to the edges in $E_j$.
    \item
    For each $i\geq 2$, no internal vertex of the path formed by $E_i$ belongs to
    $\bigcup_{j<i}V(E_j)$.
\end{enumerate}
A path consisting of a single edge is allowed.

The computational problem asks for an edge-indexed array representing a function
\[
    \mathrm{col}:E\longrightarrow\{1,\ldots,q\},
\]
where $\mathrm{col}(e)=i$ if and only if $e\in E_i$.
The algorithm may output any valid open ear decomposition.
\end{definition}

\begin{framed}
\small
\setlength{\parindent}{0pt}
\algcaption{Constructing the graph for the parity reduction}
\label{alg:parity-to-oed}\par
\smallskip
\textbf{Input:}
Bits \(x_1,\ldots,x_k\in\{0,1\}\), where \(k\ge1\),
stored in separate shared-memory cells.\par
\textbf{Output:}
A simple undirected biconnected graph \(G=(V,E)\) with
\(4k+6\) vertices and \(4k+8\) edges.\par
\smallskip
\textbf{Construction
(\textsc{ParityToOED}\((x_1,\ldots,x_k)\)):}\par
\smallskip
\begin{enumerate}[label=\arabic*.,leftmargin=1.5em,topsep=3pt,itemsep=3pt,parsep=0pt]
\item
Create two vertices \(s,t\) and, for every \(i\in\{0,\ldots,k\}\) and \(a,b\in\{0,1\}\), create a vertex \(v_{i,a,b}\).

The index \(i\) specifies the layer. Each layer has two pairs of vertices, indexed by \(a\), and the two vertices within a pair are indexed by \(b\).

Allocate an array \(A[1,\ldots,4k+8]\) to store the edges.

\item
For every \(a,b\in\{0,1\}\), in parallel, connect \(s\) to the corresponding vertex in layer \(0\) by setting
\[
    A[2a+b+1]\leftarrow \{s,v_{0,a,b}\}.
\]

\item
For every \(i\in\{1,\ldots,k\}\) and \(a,b\in\{0,1\}\), in parallel, set
\[
    A[4i+2a+b+1]
    \leftarrow
    \{v_{i-1,a,b},v_{i,a,b\oplus x_i}\}.
\]
Here \(\oplus\) denotes XOR of two bits. If \(x_i=0\), the connection preserves the index \(b\). If \(x_i=1\), the connection changes \(b\) to \(1-b\). The pair index \(a\) is always preserved.

\item
For every \(a,b\in\{0,1\}\), in parallel, connect the corresponding vertex in layer \(k\) to \(t\) by setting
\[
    A[4k+2a+b+5]\leftarrow \{v_{k,a,b},t\}.
\]

\item
Return \(G=(V,E)\), where
\[
    V=\{s,t\}\cup
    \{v_{i,a,b}:0\le i\le k,\ a,b\in\{0,1\}\},
\]
and \(E\) is the edge set stored in \(A\).
\end{enumerate}
\end{framed}

\begin{theorem}[Parity reduces to open ear decomposition]
\label{thm:parity-to-oed}
Suppose a deterministic ARBITRARY CRCW PRAM algorithm
computes an open ear decomposition, represented by
$\mathrm{col}$ as in
Definition~\ref{def:open-ear-decomposition}, for every
$n$-vertex, $m$-edge biconnected graph in parallel time
$T(n,m)$ using $P(n,m)$ processors. Then parity on $k$ input bits can be computed on the same model in parallel time
\[
    T(4k+6,4k+8)+O(1)
\]
using
\[
    O\bigl(P(4k+6,4k+8)+k\bigr)
\]
processors.

Consequently, with polynomially many processors, open ear decomposition requires
\[
    \Omega\!\left(\frac{\log n}{\log\log n}\right)
\]
worst-case parallel time on this family of simple biconnected graphs, even though every graph has $m=n+2$ edges and every open ear decomposition has exactly three ears.
\end{theorem}

\begin{proof}
Apply Algorithm~\ref{alg:parity-to-oed} to the input $x_1,\ldots,x_k$. We first describe the resulting graph, then show how to recover parity from any valid open ear decomposition.

For the analysis, let
\[
    p_0=0,\qquad
    p_i=x_1\oplus\cdots\oplus x_i
    \quad (1\leq i\leq k).
\]

For each $a,b\in\{0,1\}$, the graph contains the path
\[
    Q_{a,b}
    =
    \bigl(
        s,\,
        v_{0,a,b},\,
        v_{1,a,b\oplus p_1},\,
        \ldots,\,
        v_{k,a,b\oplus p_k},\,
        t
    \bigr).
\]
Indeed, the edge between layers $i-1$ and $i$
changes the last index by $x_i$, and
$p_i=p_{i-1}\oplus x_i$.

For each layer $i$, the map
$(a,b)\mapsto(a,b\oplus p_i)$ is a permutation
of $\{0,1\}^2$. Thus the four paths $Q_{a,b}$
are internally vertex-disjoint and together contain
every edge of $G$. Each path has $k+1$ internal
vertices and $k+2$ edges. Therefore,
\[
    |V|=2+4(k+1)=4k+6,
    \qquad
    |E|=4(k+2)=4k+8.
\]

The graph is simple and biconnected.
To see biconnectivity, deleting $s$ leaves all
remaining vertices connected through $t$, and
deleting $t$ leaves them connected through $s$.
If an internal vertex of one path is deleted, the
other three paths still connect $s$ to $t$.
The remaining portions of the affected path attach
to $s$ or $t$, so the remaining graph is connected.

Now consider any open ear decomposition
$E_1,\ldots,E_q$ returned by the assumed algorithm.
Every vertex other than $s$ and $t$ has degree two.
Consequently, any simple cycle in $G$ consists of
exactly two complete paths among the four paths
$Q_{a,b}$. In particular, the first ear $E_1$
consists of two such paths.

The unused edges now form the other two complete
$s$-$t$ paths. Each of these paths meets the vertices
of $E_1$ only at $s$ and $t$. Hence the next ear
must use one entire remaining path: a proper
nonempty segment would have an endpoint that has
not appeared in an earlier ear. The final remaining
path must then form the third ear. Thus every valid
decomposition has exactly three ears: a cycle
formed by two paths, followed by the other two
paths in either order.

We recover $p_k$ by reading only the labels of the
four edges incident to $s$ and the four edges
incident to $t$. Write
\[
    e^s_{a,b}=\{s,v_{0,a,b}\},
    \qquad
    e^t_{a,b}=\{v_{k,a,b},t\}.
\]
Among the four labels $\mathrm{col}(e^s_{a,b})$,
the label of the first ear occurs twice, while
each of the other two labels occurs once.
Choose any pair $(a,b)$ whose label occurs once.

The edge $e^s_{a,b}$ belongs to an ear consisting
of the complete path $Q_{a,b}$. There is therefore
exactly one edge incident to $t$ with the same
label. Let it be $e^t_{a',b'}$. The description of
$Q_{a,b}$ gives
\[
    a'=a,
    \qquad
    b'=b\oplus p_k.
\]
Return $b\oplus b'$. This equals
\[
    p_k=x_1\oplus\cdots\oplus x_k,
\]
as required. The argument holds for every valid
open ear decomposition returned by the algorithm.

Finally, the graph construction takes $O(1)$
parallel time with $O(k)$ processors and $O(k)$
work. Each edge has a predetermined array position,
and each edge between consecutive layers depends
on just one input bit. No two processors need to
write to the same array position. Decoding requires
only a constant number of label comparisons, so it
takes $O(1)$ time and work. This proves the stated
time and processor bounds.

If $P(n,m)$ is polynomially bounded, the resulting
parity algorithm also uses polynomially many
processors. Theorem~\ref{thm:lower-bound-of-parity}
therefore implies
\[
    T(4k+6,4k+8)+O(1)
    =
    \Omega\!\left(\frac{\log k}{\log\log k}\right).
\]
Since $n=4k+6$, this yields the claimed lower bound.

The decoding uses only equality tests on ear labels.
Thus the same reduction also works when the output
provides arbitrary distinct labels for the ears,
without specifying their order.
\end{proof}

\section{Recovering parent pointers from preparation records}
\label{sec:history}
Only the preparation phase selects original edges before witness
recovery. Its connection trees are stars, so their directions can be
recovered locally: retain every leaf-to-center direction except the
edge from the component containing the designated root.
That one edge, if present, is reversed. Processing the phases backwards
then supplies the designated roots for the next phase.
No ancestor table or LCA query is needed for this recovery.

\subsection{Recording the star contractions}
\label{sec:compact-records}
Store each original edge \(e=\{a,b\}\) as two immutable arc records
\(2e\) and \(2e+1\), with endpoints \((a,b)\) and \((b,a)\).
The edge identifier of arc \(j\) is \(\lfloor j/2\rfloor\).
At preparation boundary \(i\), the shared row \(\owner_i[v]\)
identifies the component containing original vertex \(v\), with
\(\owner_0[v]=v\).
Each component retains its center's original identifier.

When an old representative \(u\) links to a center in phase \(i\),
save the selected immutable arc identifier in \(\mathsf{sel}[u]\).
Its original tail belongs to \(u\)'s component.
After the phase's shortcut, let \(z_i[u]\) be the new representative
of old component \(u\). Write
\[
 \owner_{i+1}[v]=z_i[\owner_i[v]]
\]
for all original vertices in parallel.
Each representative retires at most once, so \(\mathsf{sel}\)
needs one entry per vertex.
The rows and selected arcs use \(nR+n\) shared words for
\(R=R_0=O(\log\log n)\) phases.
Store the owner rows as a shared table and \(\mathsf{sel}\) as a
shared vertex-indexed array. Both support constant-time direct access.

\subsection{Reverse orientation and root propagation}
\label{sec:decode-history}
For a new component \(C\), let \(\rho_{\rm next}[C]\) be its
designated original root.
The old component containing it is
\[
 x[C]=\owner_i[\rho_{\rm next}[C]].
\]
The old center has identifier \(C\).
If \(x[C]=C\), all old leaf components keep their outgoing
directions to \(C\). Otherwise the old center points to \(x[C]\),
using the edge selected by \(x[C]\) in reverse; all other leaves
still point to \(C\).
For an oriented connecting edge \(a\to b\), its tail \(a\) becomes
the designated root of the old component it leaves.
The target component \(x[C]\) inherits \(\rho_{\rm next}[C]\).
The shared buffers \(\rho_{\rm next}\) and \(\rho_{\rm cur}\) store
the designated roots at consecutive preparation boundaries.

\noindent\begin{minipage}{\linewidth}
\begin{framed}
\small
\setlength{\parindent}{0pt}
\algcaption{Parent Pointer Recovery}
\label{alg:restore}\par
\smallskip
\textbf{Input:} Preparation owner rows
\(\owner_1,\ldots,\owner_R\), selected arcs \(\mathsf{sel}\),
immutable input arcs, and designated roots \(\rho_{\rm final}[C]\)
(default \(C\)).\par
\textbf{Output:} Original-edge parents \(\Par\) and parent-edge
identifiers \(t\).\par
\smallskip
\begin{enumerate}[label=\arabic*.,ref=\arabic*,leftmargin=1.5em,topsep=3pt,itemsep=3pt,parsep=0pt]
\item \label{step:restore-root}
Set \(t[v]\gets\nil\) for all original vertices.
For each final component \(C\), set
\(\rho_{\rm next}[C]\gets\rho_{\rm final}[C]\) and
\(\Par[\rho_{\rm final}[C]]\gets\rho_{\rm final}[C]\).

\item
For \(i=R-1,R-2,\ldots,0\):
\begin{enumerate}[label=\alph*.,ref=\arabic{enumi}.\alph*,leftmargin=1.4em,topsep=2pt,itemsep=2pt,parsep=0pt]
\item \label{step:restore-target}
For each new representative \(C\), set
\(x[C]\gets\owner_i[\rho_{\rm next}[C]]\).

\item \label{step:restore-attach}
For each old representative \(u\), read
\(C\gets\owner_{i+1}[u]\); skip \(u=x[C]\).
Set \(w\gets x[C]\) when \(u=C\), and
\(w\gets u\) otherwise.
Read \(j\gets\mathsf{sel}[w]\) and its original endpoints
\((a,b)\); exchange \(a,b\) if \(\owner_i[a]\ne u\).
Set \(\Par[a]\gets b\), \(t[a]\gets\lfloor j/2\rfloor\), and
\(\rho_{\rm cur}[u]\gets a\).

\item \label{step:restore-inherit}
For each new representative \(C\), set
\(\rho_{\rm cur}[x[C]]\gets\rho_{\rm next}[C]\).

\item
Exchange the roles of the two shared root buffers.
\end{enumerate}

\item Return \(\Par,t\).
\end{enumerate}
\end{framed}
\end{minipage}\par\smallskip

Old representatives are the fixed points of \(\owner_i\);
new representatives are the fixed points of \(\owner_{i+1}\).
Processors check these conditions in constant time.
The array \(x\) and the root buffers are
shared vertex-indexed arrays with address equal to their base plus
the representative identifier.

\subsection{Correctness and cost}
\label{sec:restore-analysis}
The designated roots allow the preparation forest to attach to the
base-edge forest recovered in \S\ref{sec:additive-sf}.
In \algstep{alg:additive-sf}{step:sf-original-parents}, run recovery
in temporary arrays \(\Par_0,t_0\).
Copy their internal parents for all vertices other than the
designated roots. At each designated root, install its outgoing
base-edge parent if one exists, and keep its self-parent otherwise.
Thus internal recovery never overwrites an outgoing base edge.

\begin{lemma}[Directions in a preparation star]
\label{lem:restore-phase}
The edges selected in a preparation phase, oriented toward a
designated old component \(x\), retain every leaf-to-center direction
except \(x\to C\), which is reversed when \(x\ne C\).
\end{lemma}
\begin{proof}
The unique path from any leaf other than \(x\) to \(x\) first visits
the center \(C\). If \(x=C\), that path ends there.
Otherwise its last edge is \(C\to x\).
These are precisely all edges of the star with the stated directions.
\end{proof}

\begin{lemma}[Recovered edges point toward the designated root]
\label{lem:restore-orientation}
The recovery orients each selected original edge along its unique
path to its final designated root, with exclusive parent writes.
\end{lemma}
\begin{proof}
Induct backwards over the preparation phases.
At boundary \(i+1\), each component \(C\) has a designated vertex
\(\rho_{\rm next}[C]\). Previously oriented external edges leave
\(C\) only at that vertex; no other parent in \(C\) has been assigned
by those external edges. This holds at initialization.

The target published in \algstep{alg:restore}{step:restore-target}
is the unique old component containing that vertex.
Lemma~\ref{lem:restore-phase} gives the required direction of every
connecting edge. In \algstep{alg:restore}{step:restore-attach},
an old leaf \(u\ne x[C]\) uses its own selected arc.
The old center \(u=C\), when different from \(x[C]\), uses
\(\mathsf{sel}[x[C]]\) in reverse.
The owner test orients the immutable endpoints so that \(a\)
belongs to \(u\)'s component and \(b\) to the next component
on the path toward \(x[C]\).

The target component inherits the designated vertex.
Every other component acquires \(a\) as its designated vertex and
\(a\to b\) as its outgoing parent. It previously contained no assigned
external parent, so this write overwrites none.
Distinct old components have disjoint vertices, making all new parent
writes exclusive. This proves the induction invariant at boundary \(i\).
At boundary zero, each component is a singleton, so every selected
edge has its direction toward the final root.
\end{proof}

\begin{theorem}[Preparation recovery]
\label{thm:restore}
Algorithm~\ref{alg:restore} recovers exactly the selected preparation
forest, rooted at its designated vertices.
It takes \(O(R+1)=O(\log\log n)\) time with \(O(n)\) processors
and constant private memory. Persistent records use \(nR+n\)
shared words; temporary recovery arrays use \(O(n)\) words.
\end{theorem}
\begin{proof}
Lemma~\ref{lem:restore-orientation} proves the parent directions.
Each non-target old component contributes its one edge toward the
target in the rerooted star, so every selected edge is used exactly
once and no new edge is introduced.
Each reverse phase makes a constant number of direct shared-memory
accesses per vertex. Target publication, edge orientation, and
root propagation each take constant time with one processor per
vertex. Summing over \(R\) phases gives the time bound.
Two root buffers, the target array, and the parent outputs
use \(O(n)\) shared words. The stored owner rows and selected arcs
give the stated persistent space.
\end{proof}

\newpage
\addcontentsline{toc}{section}{References}

\bibliographystyle{alpha-three}
\bibliography{main}

\end{document}